\documentclass[aps,prd,twocolumn,superscriptaddress,nofootinbib,
eqsecnum]{revtex4-1} 
\usepackage{amsmath} %\usepackage{amssymbol}
\usepackage{amsfonts} 
\usepackage{epsfig} 
\usepackage{graphicx}
\usepackage{hyperref}

\newcommand{\beq}{\begin{equation}}
\newcommand{\eeq}{\end{equation}}
\newcommand{\bea}{\begin{eqnarray}}
\newcommand{\eea}{\end{eqnarray}}
\newcommand{\nn}{\nonumber\\}

\def\dcal{{\cal D}}

\def\hcal{{\cal H}}

\def\lcal{{\cal L}}
\def\ocal{{\cal O}}
\def\rcal{{\cal R}}

\def\pa{\partial}
\newcommand{\vev}[1]{\!\left\langle#1\right\rangle\!}
\newcommand{\eqn}[1]{Eq.~{\hspace{-2pt}}(\ref{#1})}
\newcommand{\eqns}[2]{Eqs.~{\hspace{-2pt}}(\ref{#1}),\,(\ref{#2})}

\newcommand{\secn}[1]{Sec.~\hspace{-2pt}\ref{#1}}

\newcommand{\reference}[1]{Ref.~\hspace{-2pt}\cite{#1}}

\newcommand{\half}{\frac{1}{2}}

\newcommand{\wt}[1]{\widetilde{#1}}
\newcommand{\wh}[1]{\widehat{#1}}
\def\Tr{{\rm Tr }}

\def\mn{{\mu\nu}}
\def\kl{{\kappa\lambda}}
\def\xip{{\xi'}}

\def\ie{{\it i.e., }}
\def\viz{{\it viz., }}
\def\eg{{\it e.g., }}
\def\vsig{{\varSigma}}
\def\vdelvsig{{\varDelta\varSigma}}
\def\su51{{SU(5){\otimes}U(1)}}
 
\let\oldsqrt\sqrt
\def\sqrt{\mathpalette\DHLhksqrt}
\def\DHLhksqrt#1#2{%
\setbox0=\hbox{$#1\oldsqrt{#2\,}$}\dimen0=\ht0
\advance\dimen0-0.2\ht0
\setbox2=\hbox{\vrule height\ht0 depth -\dimen0}%
{\box0\lower0.4pt\box2}}

\begin{document}

%{\hskip0.25in\mbox{\today} \hskip1in %\mbox{{\bf Draft \jobname}} 
%\hskip1in \mbox{LTH~1140,~NSF-ITP-17-135,~MCTP-17-21}}\\

\title{Renormalizable, asymptotically free gravity without ghosts or tachyons}

\author{Martin B. Einhorn$^{1,2,*}$,
D.~R.~Timothy~Jones$^{1,3,\dag}$\\
$^{1}$Kavli Institute for Theoretical Physics, Kohn Hall,\\
University of California,
Santa Barbara, CA 93106-4030\\
$^{2}$Michigan Center for
Theoretical Physics,
University of Michigan,
Ann Arbor, MI 48109
$^{3}$Dept. of Mathematical Sciences,
University of Liverpool, Liverpool L69 3BX, U.K.
}\renewcommand{\thefootnote}{\fnsymbol{footnote}}
\footnotetext{meinhorn@umich.edu}
\footnotetext{drtj@liverpool.ac.uk}
\renewcommand{\thefootnote}{\arabic{footnote}}
\setcounter{footnote}{0}

\begin{abstract}

We analyse scale invariant quadratic quantum gravity incorporating non-minimal 
coupling to a multiplet of scalar fields in a gauge theory, with particular 
emphasis on the consequences for its interpretation resulting from a 
transformation from the Jordan frame to the Einstein frame. The result is the 
natural emergence of a de~Sitter space solution which, depending the gauge 
theory and region of parameter space chosen, can be free of ghosts and 
tachyons, and completely asymptotically free. In the case of an SO(10) model, 
we present a detailed account of the spontaneous symmetry breaking, and we 
calculate the leading (two-loop) contribution to the dilaton mass.

\end{abstract}

\maketitle
\tableofcontents

\section{Introduction}
\label{sec:intro}
There has been increasing interest in the past few years in finding 
alternatives to the common lore concerning the fundamental interactions.
With no sign of supersymmetric particle production (as yet) at the LHC,
the idea that weak-scale supersymmetry (SUSY) might be the solution to the hierarchy
problem of the Standard Model (SM) has become less attractive. Secondly,
especially since it seems that our Universe may well have a positive
cosmological constant, the relationship of string theory to cosmology
seems ever more remote. The landscape of string theory vacua has
difficulty accommodating de~Sitter-like backgrounds. Further, since the
asymptotic behavior of such spacetimes is not flat, there is no
S-matrix. Motivated by these observations, in a series of recent 
papers~\cite{Einhorn:2014gfa, Jones:2015son, Einhorn:2015lzy, 
Einhorn:2016mws,Einhorn:2016fmb}, we have explored the properties of a 
renormalizable~\cite{Stelle:1976gc},
asymptotically free~\cite{Fradkin:1981hx}, classically scale-invariant,
quantum field theory (QFT) of gravity, including
matter fields in such a way that \emph{all}  couplings remain 
asymptotically free (AF).  Asymptotic freedom allows one to entertain
the possibility that this is an  ultraviolet (UV) completion of gravity
and that there is no new physics to be discovered  at higher scales. It
also allows one to make perturbative, controllable calculations at 
arbitrarily high-energy scales\footnote{Such models have been termed 
``totally asymptotically free" in \reference{Giudice:2014tma}. The notion of AF 
is distinct from nonperturbative ``asymptotic  safety''~\cite{Weinberg:1976xy}, 
which has undergone a resurgence in recent years; see e.g.   
\reference{Litim:2014uca}.}.
% Reuter:2012id, Demmel:2014hla}.}.  
Even though the QFTs we study are not truly scale
invariant because of the  conformal anomaly, it is attractive to assume
that the models are classically  scale invariant since such theories are
technically  natural~\cite{Bardeen:1995kv} in the sense that it is not
necessary to fine-tune power-law divergent loop corrections in order to
stabilize their scalar mass spectra.  Under these circumstances, all masses,
including the Planck mass $M_P$ and the cosmological constant
$\Lambda,$ arise via dimensional transmutation 
(DT)~\cite{Coleman:1973jx}. (Such a program was already proposed 
in \reference{Fradkin:1981hx}.)

Contrary to the widespread belief that renormalizable gravity violates
unitarity, having both a spin-two ghost as well as a spin-zero tachyon
in flat background, we claim that, in a de~Sitter (dS) background,
these models have no unstable fluctuations for a certain range of
couplings. (This was already known for the theory without
matter~\cite{Avramidi:1986mj}. See \secn{sec:jordanreal}.).  
There remain five zero modes which, we have argued~\cite{Einhorn:2016fmb},  
correspond to collective modes that are unphysical and, similar to gauge  
modes, do not contribute to on-shell observables. Thus, although these zero  
modes are a generic feature of all such models in a dS background, they are 
not a barrier to stability.
Our assertion is limited to quadratic order in the fluctuations,
the same order at which claims of instabilities and ghosts have been
made.  We do not know whether, in higher order when interactions are
included, this will remain true. This is closely related to the question
of unitarity, since we do not have a canonical action or a Hamiltonian
that guarantee unitary evolution.

In previous work~\cite{Einhorn:2014gfa, Jones:2015son, Einhorn:2015lzy, 
Einhorn:2016mws}, we have displayed models exhibiting DT for a range of 
couplings, within which there is a
subset of values such that the extrema are local minima of the Euclidean
action. We have also satisfied the constraints on the couplings so that
the Euclidean path integral (EPI) is convergent for all values of the
fields. We found that these minima lie within the basis of attraction of
the AF fixed point gauge model with a ``Higgs" field in the adjoint 
representation, for a certain fermion content~\cite{Einhorn:2016mws}. So
far,  we have only described the spectrum of this model qualitatively.
In this paper,  we wish to discuss the physics of this model near or
below the scale of  symmetry breaking. In the process, we shall also
substantiate our claim that the  fluctuations are stable. For this
purpose, as is often the case, it will  prove useful to pass from the
Jordan frame to the Einstein frame.

To set the stage and review our conventions, we begin with the action for gravity 
without matter. The action for renormalizable gravity can be written in several 
different equivalent forms; we take it (in the Jordan frame) as
\beq\label{eq:Jsho}
S_{ho}^{(J)}=\int d^4x\sqrt{g_J}\left[ \frac{C^2}{2a}+\frac{R^2}{3b}+c\,G 
\right],
\eeq
where $C_{\kl\mn}$ is the Weyl tensor, $R$ is the Ricci scalar, and 
$G$ is the so-called Gauss-Bonnet (G-B) term, $G\equiv C^2-2W,$ where 
$W\equiv R_\mn^2-R^2/3.$ We shall work with the Euclidean form of the 
metric with the convention for the Ricci tensor $R_\mn$ in which $R>0$ 
corresponds to positive curvature.\footnote{Because the variation of $G$ 
vanishes, the term $C^2$ can be replaced by $2W$ in \eqn{eq:Jsho}.
This often simplifies some tensor algebra.} To this must be added a 
matter action, which will be discussed in due course. Euclidean dS 
space is the $S^4$ sphere. This may be regarded as a submanifold of flat, 
Euclidean space in five-dimensions. From this perspective, the radius of the 
$S^4$ sphere is $r_0=\sqrt{12/R_0},$ where $R_0$ is the value of the 
Ricci scalar on-shell. 

The metric $g_\mn$ of \eqn{eq:Jsho}, in transverse-traceless (TT) gauges, 
includes five on-shell tensor modes as well as a scalar mode (dilaton), plus an 
additional four modes that are gauge-dependent. So this may be thought of as a 
scalar-tensor theory of gravity. One may add an Einstein-Hilbert (E-H) term $-M_P^2 
R/2$ as well as a cosmological constant $M_P^2\Lambda.$ Since they are 
UV-irrelevant, their presence does not affect renormalizability or AF, although issues 
of fine-tuning may re-emerge, at least in non-supersymmetric models. This is 
what has usually been done in the past, but, in the scenarios that we have 
described in which scalar matter is added in a classically scale-invariant 
fashion, such terms are not needed so long as DT occurs at a scale where the 
dimensionless couplings are sufficiently small that perturbative calculations 
remain reliable.

Over the years, there have been numerous papers involving higher-derivative 
gravity in a similar spirit to ours, some of which attempt to provide a complete QFT of 
gravity~\cite{Fradkin:1981hx, 
%Fradkin:1981iu, 
Barth:1983hb, Giudice:2014tma, Holdom:2016xfn}, 
possibly conformal and/or supersymmetric~\cite{Fradkin:1983tg, Fradkin:1985am}, 
while others attempt to generate the Planck mass dynamically along the lines of 
induced gravity~\cite{Adler:1982ri, 
%Buchbinder:1986wk, Odintsov:1991nd, Shapiro:1994yt, Cognola:1998ve, 
Salvio:2014soa, 
%Kannike:2015apa, 
Salvio:2017qkx, GarciaBellido:2011de}. 
This subject has been reviewed in \reference{Fradkin:1985am, 
Buchbinder:1992rb}.
These references are just a sample, and, given the extensive literature about 
higher-derivative gravity, spanning more than 50 years, we shall have to limit 
further citations to those that are of immediate relevance. 

An outline of the subsequent sections is as follows: In the next
section, we  discuss aspects of the stability of the model in de~Sitter
background, the  controversy over the sign of $b,$ and some of the
difficulties establishing that  QFTs of this sort are (or are not)
unitary. In \secn{sec:jordanreal}, we review the  addition of matter in
the Jordan frame, taking up the simplest example of the  real scalar
field, while in \secn{sec:EFreal}, we transform the same model to the 
Einstein frame in order to elaborate on several points not discussed in
our  previous papers. Then in \secn{sec:nonabelian}, we apply these
methods to the  case of the $SO(10)$-model, which is a prototype for any
such non-Abelian  gauge theory coupled to gravity. In \secn{sec:ssb},
we turn to the issue of  spontaneous symmetry breaking (SSB) in this
model, emphasizing the  differences from a similar calculation in the
Jordan frame. Then we embark  upon a discussion of the resulting
particle spectrum in this model for the vector  bosons
(\secn{sec:vbmass}), the heavy scalars (\secn{sec:scalarmass}), the 
curvature fluctuations (\secn{sec:rho}) and finally the dilaton mass 
(\secn{sec:dilaton}) arising at two-loop order.  Following some remarks
on the resulting low-energy effective field theory (\secn{sec:left}),
we summarize our  results and discuss open questions in
\secn{sec:conclude}. There follow two  appendices with details useful
in the body of the text.  In  Appendix~\ref{sec:conftransf}, we review
how the curvature tensor transforms  under conformal transformations,
and in Appendix~\ref{sec:algebra} our Lie algebra notation
and the form of the model after SSB to  $SU(5){\otimes}U(1).$

\section{Stability, asymptotic freedom, and unitarity}\label{sec:unitarity}
 
Everyone who has considered renormalizable gravity agrees that $a>0$ is 
necessary and sufficient for this coupling to be AF. As we have previously
mentioned~\cite{Einhorn:2016mws}, the appropriate sign of $b$ has been
subject to some dispute, and we shall take up this issue below.
 
We adopt the assumptions of Euclidean quantum gravity~\cite{Hawking:1978jn} to the 
extent that they are known. To some extent, these have been reviewed in 
\reference{Fradkin:1985am, Buchbinder:1992rb}. Our philosophy is very close
to that elaborated by Christensen and Duff~\cite{Christensen:1979iy}
and by Avramidi~\cite{Avramidi:1986mj}\,\footnote{These will be  further 
reviewed below. \reference{Christensen:1979iy} does not consider  
renormalizable gravity, and \reference{Avramidi:1986mj}  mentions
the inclusion of matter only in passing.}. A basic tenet of this
approach is that the Euclidean path integral (EPI) be  convergent for
all values of the fields. Unlike E-H gravity, 
integrating over conformal modes presents no special difficulties.  This
requires both $a$ and $b$ in \eqn{eq:Jsho} to be positive for
sufficient large scales where the ``classical" approximation is  valid.
This appears to be a minimal requirement for  the existence of candidates
for stable ``vacuum" states in QFT. In flat  spacetime, the requirement
that the Euclidean action be bounded below together with certain 
others~\cite{Osterwalder:1973dx}, eventually 
allows for analytic continuation to Lorentzian signature with an action 
that respects CPT invariance and unitarity. Whether something similar 
is true for the extension of gravity given in \eqn{eq:Jsho} is not
known.  It should not be difficult to extend reflection-positivity to
Euclidean  renormalizable gravity, but cluster decomposition obviously
must be modified  for a compact spacetime such as $S^4.$ For Euclidean
spacetimes without  boundaries, this would imply that there are not
degenerate no-particle states. In  particular, apparently degenerate
no-particle states must have finite tunnelling  amplitudes between them
so that they can be superposed. For example, this is  familiar in flat
space when there are finite action solutions of the classical  equations
of motion (EoM) for Euclidean signature (instantons). In that case, 
there are degenerate no-particle states in perturbation theory for
which, as a  result of non-perturbative affects, the degeneracy is
removed. 

There exist persistent doubts about unitarity in this class of theories.
Unitarity is certainly suspect in theories with actions containing 
both quadratic curvature terms of the kind exhibited in 
\eqn{eq:Jsho}\ \emph{and} an explicit linear term $-M_P^2 R$, 
because of the following observations, which were 
raised originally in \reference{Stelle:1976gc}. In the presence of a 
non-zero Planck mass $M_P,$ the propagator in flat space contains a term in 
the tensor mode that behaves as 
\beq\label{eq:ghost}
\frac{1}{q^2(q^2+M_P^2)}=\frac{1}{M_P^2}
\left(\frac{1}{q^2} - \frac{1}{q^2+M_P^2}\right).
\eeq
Thus, if the graviton term $1/q^2$ has the usual sign, the second term 
corresponds to a massive, spin two particle with negative kinetic
energy,  i.e., a ghost. Further, in the scalar sector, there remains a
particle with  mass~\cite{Stelle:1976gc} $m_0=\sqrt{-b}M_P/2,$ where 
$M_P=1/\sqrt{8\pi G_N}$ is the so-called ``reduced" Planck mass or
string  scale. Thus, there is a tachyon instability for $b>0$  in flat
background, a primary reason some have argued that $b<0.$ Yet, as
remarked above, $b>0$ is the sign required for convergence of the EPI.
Since, however, we have demonstrated (and will confirm here) that 
the phase of our model having terms both linear and quadratic in curvature
exists only below a definite scale that is determined by DT, the argument 
based on \eqn{eq:ghost} does not apply. We shall make some further comments 
about unitarity below.

Another argument suggesting that $b<0$ would be preferable goes as follows: 
One adds to $R^2$ a term with an auxiliary field $\chi$ 
\beq\label{eq:aux}
\frac{1}{3b}R^2\mp\half\left(\! \chi^2{-}\frac{\xi}{2}R\!\right)^{\!2}\!\!{=}
\!\left[\frac{1}{3b}\mp\frac{\xi^2}{8}\right]\!R^2\pm
\frac{\xi \chi^2}{2}R\,\mp\,\frac{\chi^4}{2}.\!
\eeq
with $\xi$ an arbitrary ``coupling constant." The sign of $\xi$ must be chosen 
to be the same as the sign of $\vev{R},$ so that $\vev{\chi}^2\!=\xi \vev{R\,}/2$ has a solution 
for real $\vev{\chi}.$ The sign of the added term must be chosen to be opposite to the 
sign of $b,$ so that the coefficient of $R^2$ on the RHS can be taken to vanish 
$(\xi^2=8/|3b|).$ Thus, it seems that the original term in the Lagrangian density 
proportional to $R^2$ is equivalent to a non-minimal gravitational coupling of a scalar 
field together with its self-interaction. We then see that if $b<0,$ the linear term in $R$ 
corresponds to attractive gravity, and the ``potential term" $\chi^4$ is bounded 
below. This is frequently used~\cite{Sotiriou:2008rp} 
%,DeFelice:2010aj, Nojiri:2010wj, Narain:2016sgk} 
to argue that the sign demanded physically is $b<0.$ This sign is the opposite of that 
required for convergence of the EPI and for AF of $b.$

We {\it do not}, however, subscribe to this popular belief that $b(\mu)<0$ (for 
sufficiently large scales $\mu$) because the field $\chi,$ unlike an independent 
dynamical degree of freedom (DoF), is inextricably linked to the scalar curvature, \ie
$\chi^2=\xi R/2.$ From the point of view of the EPI, the preceding construction 
is misleading; one cannot simply add such a term and integrate over $\chi$ 
since, having insisted $b(\mu)>0$ at large scale, the integral over $\chi$ would 
diverge. To introduce an auxiliary field, one must actually add to the integrand of 
the EPI a term proportional to $\dcal\chi^2\, \delta(\chi^2-\xi R/2),$ 
or its equivalent. 

To confirm the fallacy in such arguments, consider the far simpler situation in 
ordinary $\phi^4$ field theory in flat spacetime with potential 
$V(\phi)=m^2\phi^2/2+\lambda\phi^4/4.$ 
It is generally believed that, in order to have a sensible ground state, one must 
have the renormalized coupling $\lambda(\mu)>0,$ at least for some range of 
relatively large scales\footnote{The sign of $\lambda$ is a renormalization 
group invariant since $\lambda=0$ yields free field theory. $\lambda>0$ is 
IR-free and not AF, so this must be regarded as an effective field theory.}. 
Following a procedure similar to the previous one, we write 
\beq
\frac{\lambda}{4}\phi^4\mp\frac{1}{4}\big(\sigma-\xi\phi^2\big)^2=
\frac{1}{4}\left(\lambda\mp\xi^2\right)\!\phi^4
\pm\frac{\xi\sigma}{2}\phi^2\mp \frac{1}{4}\sigma^2. 
\eeq
$\sigma$ is an auxiliary field\footnote{Note that, with this definition,
$\sigma$ has dimensions of mass-squared.} for which $\vev{\sigma}=\xi\vev{\phi}^2.$ 
To be able to cancel the $\phi^4$ term on the RHS, thereby reducing the action 
for $\phi$ from quartic to quadratic, we must choose the sign of 
the added term to be opposite to that of $\lambda.$ For the ``potential 
term" $\sigma^2$ to be bounded below, the last term must be positive. By the 
logic above, we ought then to demand $\lambda<0,$ the very opposite of what 
we required initially!

We conclude that one may not treat an auxiliary field such as $\sigma$
as if it can be taken ``off-shell" for fixed values of the other
fields on which it depends. Conversely, it may also not be consistent
to discuss the behavior of a dynamical field such as $\phi$ for
arbitrary values of the auxiliary field. The construction is also wrong
in detail, because the equation $\xi(\mu)^2=\lambda(\mu)$ is not in 
fact correct for arbitrary $\mu;$ in short, it is not renormalization
group invariant\footnote{For further discussion on this point, see,
e.g., Sec.~II of~\reference{Jack:2000nm}.} (RGI). Similarly, in the 
gravitational case, the relation $3b(\mu)=8/\xi(\mu)^2$ is not RGI. In
sum, although one may introduce an auxiliary field in the manner
outlined here, one can be misled drawing conclusions based on treating
it as an independent DoF. 

As an aside, this same issue arises in other models involving
polynomials in $R$ of even higher degree, so-called $f(R)$ models of
gravity. It seems that a similar sign error afflicts many of those
treatments in the literature.\footnote{For reviews of such models, see
e.g.\ Refs.~\cite{Sotiriou:2008rp}.
%DeFelice:2010aj, Nojiri:2010wj,  Narain:2016sgk. 
For further extensions of this method, see \reference{Rodrigues:2011zi}. More
recently, Narain~\cite{Narain:2017tvp} has argued that there is a
conflict between the Lorentzian and Euclidean formulations. Our
expectation would be that, once again, this is a reflection of a similar 
sign issue. (See \hyperref[note:added]{{\bf Note Added}~[\ref{note:added}.])}\label{footnote:reviews}}

In this paper, we shall assume that the cosmological constant of the effective 
field theory at low energy is positive, so we shall only be concerned with 
de~Sitter-like solutions of the model. That assumption happens to be correct 
in the classically scale-invariant theories we have studied, although we have 
not investigated whether it must be true in all such theories.

An effort similar to ours embracing a classically scale-invariant action
for both matter and gravity has been called 
``Agravity"~\cite{Salvio:2014soa}. However, our approach is
fundamentally different inasmuch as these authors insist  that $b<0$ for
the reasons reviewed above. Given that the ($b<0$) model is no longer
AF, they treated renormalizable gravity as an effective field theory. It
is an improvement over the E-H theory in the same way that the
electroweak theory is an improvement over the Fermi model and may allow
some speculations about physics beyond the Planck scale. More 
recently~\cite{Salvio:2017qkx}, by considering non-perturbative
possibilities  rather than adding new dynamical degrees of freedom, they
have speculated  that perhaps the non-AF theory is correct to infinite
energy. We prefer to explore the possibility that the AF model
$(b>0)$ is the completion of the E-H theory, that perturbation
theory continues to hold, and no new physics is required at higher
scales, which, we contend, would be a far more compelling result.

As we have indicated in past work~\cite{Einhorn:2016mws} and has been 
emphasized long ago in Refs.~\cite{Christensen:1979iy} and 
\cite{Barth:1983hb}, further 
complications and opportunities arise in the presence of a cosmological 
constant, even though the curvature may be small. In that case, flat space is not 
a solution to the EoM, so some of the foregoing issues may disappear. Our 
point of view overlaps with that adopted by Avramidi~\cite{Avramidi:1986mj},
%, Avramidi:2000bm}, 
who explicitly included $M_P$ and $\Lambda$ in his action and who 
emphasized that, so long as his couplings and masses obeyed certain 
inequalities, neither the scalar nor the tensor modes 
present instabilities. In the present notation, he 
showed that the tensor modes are stable and ghost-free for $a>0,$ 
$\Lambda>0,$ and $2/(3b)<1/a+M_P^2/(16\Lambda).$ Moreover, there is no 
instability in the scalar sector provided 
$M_P^2/(16\Lambda)<2/(3b),$ which is compatible with the tensor constraint. 
These inequalities can even be satisfied in the classically scale-invariant case 
where $M_P\to0$ (for fixed $\Lambda.$) When matter is included\footnote{For 
$b<0,$ the one-loop corrections to the effective action in dS background should 
have an imaginary part, reflecting an instability. This is another reason that 
we believe that Agravity~\cite{Salvio:2014soa} is not self-consistent.}, for the 
cases we studied~\cite{Einhorn:2016mws, Einhorn:2015lzy, Einhorn:2014gfa}, 
the inequalities were modified, but there still existed regions of parameter space 
where there were no instabilities. 

Nevertheless, in calculating the one-loop correction to dS space, there remain 
five zero modes that seem to be universally present in both Einstein gravity 
and in renormalizable gravity, with or without the inclusion of matter. 
As we have reviewed elsewhere~\cite{Einhorn:2016fmb}, these so-called 
non-isometric, conformal Killing modes have a rather long history. We have 
argued that these reflect a collective mode that, in four-dimensions, is peculiar 
to the $S^4$ manifold. If so, they will be present not just at one loop but to all orders 
in perturbation theory. However, unlike other occurrences of such coherent 
motions, we claim the corresponding collective degrees of freedom (DoF), i.e., the 
``center of mass" coordinates, are unphysical and not relevant to the 
determination of the stability of dS background. They nevertheless do enter into 
the calculation of various gauge-invariant quantities, such as the on-shell 
effective action. The essential issue is whether or not there is a collective 
coordinate missing from the effective action. We presume not.

As a result of the foregoing, we believe that there exists a renormalizable, 
theory of gravity that, when matter fields are included, can yield new models 
that (1)~undergo DT in perturbation theory, 
(2)~yield a positive cosmological constant, (3)~are locally stable for a range of 
couplings, and (4)~are AF in all couplings. So far, we have 
confirmed this for only one such model~\cite{Einhorn:2016mws}, but it is surely not unique. 
The issue of unitarity remains unresolved, but it is far more subtle than 
has been treated thus far in the present context. For example, one of the 
lessons from considering QFT in curved 
spacetime~\cite{Birrell:1982ix, Parker:2009uva} is that the so-called no-particle state 
can appear completely different to observers in different frames, resulting in the 
definition of particle states correspondingly different.

In the next section, we expand on the way in which these results have been 
achieved. In previous papers, we used the renormalization group 
to determine the one-loop effective action. This method makes some 
assumptions that direct calculations via path integrals avoid. In the next 
section, we shall discuss this in the simplest case, that of a real scalar 
field~\cite{Einhorn:2015lzy}, but most of these points apply to the non-Abelian 
case as well, as will be discussed in \secn{sec:nonabelian}.

\section{Including matter fields in the Jordan frame}\label{sec:jordanreal}

We discussed DT in pure gravity, \eqn{eq:Jsho}, in \reference{Einhorn:2014gfa}, 
and shall not repeat that here. Matter can be added in many forms, and our goal 
is to focus on non-Abelian models, in particular, on the $SO(10)$ model 
discussed in \reference{Einhorn:2016mws}. However, there are a few points 
that can be more easily stated in the simplest case, that of the real scalar 
field~\cite{Einhorn:2015lzy}. It is also easier to have the experience of 
transforming to the Einstein frame in that case, as we shall do in 
\secn{sec:EFreal}, before proceeding to the non-Abelian gauge theory
in \secn{sec:nonabelian}.
For pedagogical reasons, then, we shall first reconsider the real field, taking 
the opportunity to clarify certain points omitted from our earlier paper.

To the action in \eqn{eq:Jsho}, we add the action for a single, real field $\phi:$
\begin{subequations}\label{clJaction}
\begin{align}\label{eq:sclJ}
S_{cl}^{(J)} &\equiv S_{ho}^{(J)}+S_m^{(J)},\\
\label{eq:smatJ}
S^{(J)}_m&\equiv \int d^4x\sqrt{g_J}\left[ \half
(\nabla\phi)^2+\frac{\lambda}{4}\phi^4-\frac{\xi\phi^2}{2}R\right].
\end{align}
\end{subequations}
Defining the rescaled couplings $y\equiv \lambda/a,$ $x\equiv b/a,$ we showed 
in \reference{Einhorn:2015lzy} that this model in dS background has a 
single ultra-violet fixed point (UVFP) at $\xi=0, y=0, x\approx 39.8.$ Given that 
all couplings are AF, the classical approximation ought to be increasingly 
accurate the higher the scale. In that paper, we derived the form of the one-loop 
corrections to the effective action using RGI, the known $\beta$-functions and 
generic form of the corrections in dS background. However, this ``short-cut" 
has its limitations. It does not necessarily reveal all constraints on the couplings 
and would not produce the imaginary part present if the perturbative 
corrections were unstable. 
These can only be revealed by explicitly calculating the one-loop effective 
action. Here, we shall review that calculation via the EPI in a ``classical" 
background field given by 
$\wh{g}_\mn(x),\varphi(x).$ For our purposes, it will suffice to consider the 
corrections on mass shell, i.e., where the effective action has extrema. To 
zeroth order, i.e., classically, the on-shell values of neither $\varphi$ nor $R$ 
can be known since the classical action $S_{cl}^{(J)}$ is scale invariant; 
however, the dimensionless ratio $\phi^2/R$ can be fixed. The first-
variation of the classical action gives 
\begin{align}
\frac{\delta S_{cl}^{(J)} }{\delta\phi}= -\Box\phi+\lambda\phi^3-\xi\phi R,
\label{eq:var1phi}
\end{align}
\begin{align}
&{-}\frac{\delta S_{cl}^{(J)} }{\delta g^\mn}=\frac{1}{6a}\bigg[4RR_\mn 
{-} 12 R^\kl R_{\mu\kappa\nu\lambda}{+}
\cr
&\hskip8mm g_\mn\big(3R_\kl^2{-}R^2 \big){+}
\big( 2\nabla_\mu\nabla_\nu R {+} 
g_\mn\Box R{-}6\Box R_\mn \big)\bigg] {+}\cr 
&\hskip7.5mm \frac{2}{3b}\bigg[\frac{g_\mn}{4}R^2{-}RR_\mn{+} 
\big({\nabla}_\mu{\nabla}_\nu{-}{g}_\mn\Box\big)R\bigg]
{-}\half\bigg[T_\mn{-}\cr
&\hskip7.5mm 
\xi\phi^2\Big[R_\mn-g_\mn R\Big]{+}
\Big[{\nabla}_\mu{\nabla}_\nu-{g}_\mn\Box\Big]\frac{\xi\phi^2}{2}\bigg],
\label{eq:var1g}
\end{align}
\beq
\mathrm{where}\ T_\mn
\equiv \nabla_\mu\phi \nabla_\nu\phi
-{g}_\mn\left[\frac{1}{2} ({\nabla}\phi)^2 
{+}\frac{\lambda}{4}\phi^4\right].\nn
\eeq
\indent It is difficult to characterize the most general solution of these equations. 
Most sufficiently symmetric solutions of Einstein's equations continue to hold for 
these modified equations, such as the Schwarzschild and 
Schwarzschild-de~Sitter solutions~\cite{Nelson:2010ig}. 
We can get a hint of what may be necessary if we take the trace of 
\eqn{eq:var1g}:
\beq
-2g^\mn\frac{\delta S_{cl}^{(J)} }{\delta g^\mn}{=}-\frac{4}{b}\Box R+
(\nabla\phi)^2+\lambda\phi^4
+\xi(3\Box -R)\phi^2.
\eeq
(The terms in $1/a$ cancel out of the trace because of classical conformal 
invariance of the Weyl action.)
Writing $\Box\phi^2=2\phi\Box\phi+2(\nabla\phi)^2,$ the right-hand side 
becomes
\beq\label{eq:trvar1g}
-\frac{4}{b}\Box R+(6\xi+1)(\nabla\phi)^2+6\xi\phi\Box\phi+
\lambda\phi^4-\xi\phi^2R.
\eeq
Only for the conformal values $b\to\infty, \xi=-1/6$ does \eqn{eq:trvar1g} 
become identical to $\phi$ times \eqn{eq:var1phi}. On the other hand, there are 
solutions other than the conformal limit that are mutually compatible with the 
vanishing of both \eqns{eq:var1phi}{eq:var1g}. For example, in the case that 
$\phi$ and $R$ are constant (corresponding to Euclidean dS space,) both 
equations are satisfied when $\lambda\phi^2=\xi R.$ 

Our first goal here is to make more explicit the requirements for calculating the 
one-loop effective action. Using the standard background field method of 
quantization by the path integral\footnote{This has been summarized in the present context in an 
appendix in \reference{Einhorn:2014gfa}.}, 
we expand the classical action $S_{cl}^{(J)},$ \eqn{clJaction},
about a generic background by writing 
$\phi(x)=\varphi(x)+\delta\phi(x),$ $g_\mn\equiv \wh{g}_\mn(x)+h_\mn(x),$ 
expanding in a Taylor series about $\varphi(x),\wh{g}_\mn(x),$
and dropping the term linear in the ``quantum fields" 
$\delta\phi(x), h_\mn(x).$ The one-loop 
correction is obtained from the terms second order in the fluctuations 
$\delta\phi(x),h_\mn(x).$ The tensor $h_\mn$ can be conveniently 
decomposed in the transverse traceless (TT) gauge
\beq \label{eq:TT}
h_\mn=h^\perp_\mn{+}\wh{g}_\mn h/4{+}\ldots,
\eeq 
where $h\equiv \wh{g}^\mn h_\mn$ and $h^\perp_\mn$ is transverse 
$(\wh{\nabla}^\mu h^\perp_\mn=0)$ and traceless 
$(\wh{g}^\mn h^\perp_\mn=0).$ The other terms represented by the 
ellipses involve gauge-dependent vector and scalar modes. 
After a lengthy calculation, this procedure yields 
\begin{align}\label{eq:jaction2}
\begin{split}
\hskip-3mm S^{(2)}&{=}\wh{S}^{(0)}{+}
\half\!\int\!\! d^4x \sqrt{ \wh{g} } \Bigg[\!
\big( \wh{\nabla}\delta\phi\big)^2 
{+}\! \left(3\lambda\varphi^2\! {-}\,\xi\wh{R}\right)\!(\delta\phi)^2\!
\\& 
{-}\delta\phi\frac{3\xi\varphi}{2}
\bigg[\Delta_0\Big[\frac{\wh{R}}{3} {-}\frac{2\lambda\varphi^2}{3\xi}\Big]\bigg] h 
{+}2 \xi\varphi\,\delta\phi\wh{R}_\mn h^\perp{}^\mn{+}\!
\\&
\hskip2mm \frac{3}{8b}h\bigg[\Delta_0\Big[{-}\frac{\wh{R}}{3}\Big]
\Delta_0\Big[{-}\frac{b\xi\varphi^2}{4} \Big] \bigg]h+
\\&
 \hskip2mm \frac{1}{2a}h_\mn^\perp\bigg[\Delta_2\Big[\frac{ a\xi\varphi^2}{2}{+}
\frac{\wh{R}}{3}\Big(1{-}\frac{2a}{b}\Big)\Big]\Delta_2\Big[\frac{\wh{R}}{6}\Big] 
\bigg]h^\perp{}^\mn\!{+}
\\& 
\hskip3mm\mathit{CT}+\mathit{other}\Bigg],
\end{split}
\end{align}
where the background metric $\wh{g}_\mn$ is to be used for contractions 
and covariant derivatives. The terms represented by $\mathit{CT}$ indicate 
implicit counter\-terms necessary to render the effective action finite after 
integration over the quantum fields. Those represented by $\!\mathit{other}\/$ 
are gauge-dependent and vanish on-shell, i.e., when the background fields 
satisfy their EoM. The symbols $\Delta_j[X]\equiv -\Box_j+X$ 
for integer $j$ involve the so-called constrained Laplacian, $\Box_j,$ upon 
which we elaborate further below.

This expression is to be inserted into the EPI and the integral over the quantum 
fields $\delta\phi, h, h^\perp_\mn$ performed. For a generic background, this 
cannot be done analytically, but, analogous to the flat space effective potential, for 
$\varphi=\varphi_0$ and $\wh{R}_\mn=R_0\, \wh{g}_\mn/4$ with $\varphi_0, R_0$ 
constant, the integral can be carried out. If we further require that the 
background be on-shell, 
$\lambda\varphi_0^2=\xi R_0,$ the quadratic action for the fluctuations can be 
put into the form 
\beq\label{eq:jaction3}
\begin{split}
\delta^{(2)}S^{(2)}_{os}=\half\!&\int\! d^4x \sqrt{ \wh{g} } \Bigg[\delta\phi
\Delta_0\Big[2\xi R_0\Big]\delta\phi-
\\& 
\delta\phi\frac{3\xi}{2}\sqrt{\frac{\xi R_0}{\lambda}}
\Delta_0\Big[{-}\frac{R_0}{3}\Big] h +
\\&  
\frac{3}{8b}h\bigg[\Delta_0\Big[{-}\frac{R_0}{3}\Big] 
\Delta_0\Big[{-}\frac{b\xi^2R_0}{4\lambda} \Big] \bigg]h+\\
&\hskip-19mm 
\frac{1}{2a}h_\mn^\perp\!\bigg[\Delta_2\Big[\frac{ a\xi^2R_0}{2\lambda}{+}
\frac{R_0}{3}\Big(1{-}\frac{2a}{b}\Big)\!\Big]\Delta_2\Big[\frac{R_0}{6}\Big] 
\bigg]h^\perp{}^\mn\Bigg],\!
\end{split}
\eeq
where the $\mathit{CT}$ and $\mathit{other}$ terms have been
suppressed. We take the background to be the sphere $S^4$ with curvature
$R_0.$ In flat five-dimensional Euclidean space, this corresponds the
four-sphere of radius $ \mathrm{r}_0\equiv\sqrt{12/R_0}$ and angular 
volume $\omega_4=8\pi^2/3.$ Thus the Euclidean spacetime volume
$V\equiv \omega_4\mathrm{r_0}\!^4=384\pi^2/R_0^2$, is finite in this
approximation. Following \reference{Fradkin:1981hx, Avramidi:1986mj},
%Avramidi:2000bm
we expand in normalized eigenfunctions of the
``constrained" Laplacian $\Box_j$ in order to determine whether the
modes are stable and to be able to deal with the mixing between
$\delta\phi$ and $h.$ 
$\Box_j=\wh{g}^\mn\wh{\nabla}_\mu\wh{\nabla}_\nu,$ where
$\wh{\nabla}_\mu$ represents the covariant derivative acting on a field
of ``spin" $j.$ For example, $\Box_0$ represents the Laplacian on the
background manifold acting on a scalar field such as $\delta\phi.$ 
$\Box_1$ represents the Laplacian acting on a conserved vector field,
$\varepsilon^\mu$ with $\wh{\nabla}_\mu \varepsilon^\mu=0.$ $\Box_2$
represents the Laplacian on the background vector bundle acting on
tensor fields such as $h^\perp_\mn,$ which is transverse and traceless.
 (Further details with references to the literature can be found 
in \reference{Fradkin:1981hx}, summarized in \reference{Avramidi:1986mj}.)
%, Avramidi:2000bm
Since the $S^4$ sphere is compact, the eigenvalues of the elliptic operator $-\Box_j$ 
are discrete and nonnegative. Explicitly, they are given by 
\begin{align}\label{eigenmodes}
\begin{split}
{-}\Box_j Y^{nj}_{\ell,m}&=
\mathrm{r}_0^{-2}\lambda_{nj}Y^{nj}_{\ell,m},\\ 
\lambda_{nj}&=n(n{+}3){-}j,\ \ n{=}j, j{+}1,\ldots.
\end{split}
\end{align}
for $n, j\geq0.$ The indices $(\ell,m)$ denote the various states of
the degenerate eigenvalue. We shall not need their precise definitions;
we just need to know the total degree of
degeneracy~\cite{Hamermesh:grpth}, $d_{nj}=(2n{+}3)(2j{+}1)
\big((n{+}1)(n{+}2)-j(j{+}1)\big)/6,\ n{\geq} j{\geq}0.$ For a scalar
field $j{=}0,$ $\lambda_{n0}=n(n{+}3)$ is simply the value of the 
quadratic Casimir of the angular momentum generators of $SO(5).$ It has 
degeneracy $d_{n0}=(2n{+}3)(n{+}1)(n{+}2)/6.$

Expanding the fluctuations in terms of the eigenfunctions $Y^{nj},$ normalized 
to one on the unit $S^4,$ 
\begin{align}\label{eq:jaction4}
\begin{split}
\frac{1}{\omega_4}\delta^{(2)}S_{os}&\!=\frac{3\xi}{ay}\sum_{n=0} d_{n0} \bigg[
2\Big[\lambda_{n0}{+}24\xi \Big]\!
\left(\!\frac{\delta\phi_n}{\varphi_0}\!\right)^{\!2}\!{-}
\\&\hskip5mm
3\xi\Big[\lambda_{n0}{-}4\Big] \frac{\delta\phi_n}{\varphi_0} h_n {+}
\\&\hskip4mm 
\frac{y}{16x\xi}\Big[\lambda_{n0}{-}4\Big] 
\left[\lambda_{n0}{-}\frac{3x\xi^2}{y} \right] h_n^2\bigg]{+}
\\&\hskip-13mm 
\frac{1}{8a}\sum_{n=2} d_{n2}
\left[\lambda_{n2}{+}\frac{6 \xi^2}{y}{+}4\left(1{-}\frac{2}{x}\right)\!\right]
\Big[\lambda_{n2}{+}2\Big] h_n^\perp{}^2,\!
\end{split}
\end{align}
where $\omega_4\equiv 8\pi^2/3,$ $y\equiv \lambda/a.$ It is the ratios
$y$ and $x$ that approach finite UVFPs. As in pure gravity, we take
$a>0$ so that it will be AF. Because the $\lambda_{nj}$ monotonically
increase with $n,$ the modes will certainly be nonnegative for $n$
sufficiently large but finite. Hence, we just need to determine whether
a finite number of modes are stable.

First, however, we must deal with the fact that each of these sums
formally diverge as $n\to\infty$ and are rendered finite by adding
renormalization counterterms that have not been explicitly included
above\footnote{Given the forms of $\lambda_{nj}$ and $d_{nj}$ above,
the zeta-function method~\cite{Hawking:1976ja} naturally comes to mind.
This involves certain subtleties in applications such as this involving
products of quadratic, elliptic operators, as reviewed in
\reference{Elizalde:1997nd}, but these should not affect our
arguments.}. Regardless of how renormalization is carried out,
instabilities at low $n$ for arbitrarily large scale $\mu$ will not be
removed. They introduce singularities for certain values of the 
couplings, which renormalization does not do and which, because of AF,
will not be removed in higher order. Zero modes, such as those
associated with $\lambda_{10}=4,$ must be subtracted and dealt with
separately and will be discussed below.

Let us begin our stability analysis with the tensor modes $h_n^\perp{}^2$ in 
\eqn{eq:jaction4}. The lowest mode has $n{=}2$ for which $\lambda_{22}=8.$ Hence, 
the factor $\lambda_{n2}{+}2$ will be positive for all $n,$ but the first factor will be 
positive only for $2{+}\xi^2/y>4/(3x).$ Given the aforementioned properties of the 
UVFP, together with the information that $\xi^2/y$ actually vanishes as the UVFP is 
approached, this inequality is easily satisfied at sufficiently high scales, so all 
the tensor modes are stable, at least at sufficiently high scale. An instability 
at lower scales would be associated with a phase transition.

What about the scalar $(j=0)$ modes? We must determine under what 
conditions the quadratic form in $(\delta\phi_n, h_n)$ is nonnegative. For 
$n=0,$ $\lambda_{00}=0,$ so this is simply
\beq\label{eq:mode00}
 \frac{9\xi^2}{a y}\bigg[
16
\left(\!\frac{\delta\phi_0}{\varphi_0}\!\right)^{\!2}+
4\frac{\delta\phi_0}{\varphi_0} h_0 +\\
\frac{1}{4}h_0^2\bigg]
\eeq
This quadratic form has one eigenvalue equal to $+585\xi^2/(4a y),$ which is 
positive since\footnote{$y>0$ is required for convergence of the EPI.} $y>0.$ Its 
eigenvector has $(\delta\phi_0/\varphi_0,h_0)\propto (8,1).$ The other 
eigenvalue is $0$ with eigenvector $(\delta\phi_0/\varphi_0,h_0)\propto (-1,8).$ 
This zero mode is the dilaton and should have been anticipated: Under 
the assumption that the background field has nonzero curvature $R_0$ and 
nonzero scalar field $\varphi_0,$ the classical scale invariance is spontaneously 
broken, so there must be a Goldstone boson. We can regard the preceding 
calculation as a purely classical determination of the eigenvalues for small 
fluctuations, so it must reflect this Goldstone mode. When we insert this into the 
EPI and integrate over the fields, this becomes the one-loop correction. In so 
doing, the zero mode must be factored out in order to obtain a finite result.
To this order, this corresponds to a flat direction of the effective potential. 

Since the scale invariance is anomalous and not a symmetry of the QFT, this 
mode can get a nonzero mass $m_d$ in higher-order. Indeed, at two-loop order, 
we argued in \reference{Einhorn:2015lzy} that $m_d^2\ne0$ and can be 
positive for some range of values of ${x,\xi,y}.$ (See \eqn{dilatonmass2} 
below.) In particular, it is positive near the UVFP, so this zero mode ultimately 
does not destroy local stability. 

The next mode is $n=1,$ for which $\lambda_{10}=4$ with degeneracy 5. 
Clearly, the quadratic form degenerates to 
 \beq
\frac{24\xi}{ay} (6\xi{+}1)(\delta\phi_1/\varphi_0)^2.
 \eeq
Since $y>0$ was required for stability of the $n=0$ mode, we must therefore 
have $\xi>0$ for stability of this mode. Obviously, its eigenvector 
$(\delta\phi_1/\varphi_0,h_1)\propto (1,0).$ The second eigenvalue is obviously 
zero due to fluctuations in the direction $(0,1).$ Thus, there are 5 zero modes 
associated with the fluctuation $h_1$ with $\delta\phi_1=0.$ These existed 
already in the pure gravity case and are present in all models with $S^4$ 
background on-shell. As mentioned earlier, we have argued in 
\reference{Einhorn:2016fmb} that these five zero modes are artifacts of the 
$SO(5)$ isometry of dS corresponding to an unphysical coherent fluctuation, a 
would-be collective mode corresponding to the motion of the center-of-mass 
coordinate of the $S^4$ sphere, so we expect these zero modes to persist 
to all orders in perturbation theory. They are not Killing vectors, but are 
conformal Killing vectors not usually associated with physical isometries of the 
action. 
They are peculiar to an $S^4$ background and  even occur for the  E-H
action~\cite{Christensen:1979iy}. We have argued that these unique modes
do not  reflect an actual physically-allowed fluctuation.  As they only
exist for an $S^4$  background, it seems likely that more realistic
models will not have such unphysical  collective coordinates. Further
research is required to determine whether some non- perturbative effect,
such as tunneling to a background with a different topology, leads to  a
different background that removes such modes.

What about the $n=2$ mode, for which $\lambda_{20}=10$? The quadratic form 
becomes 
\beq
\frac{6\xi}{ay}\!\left[
2(5+12\xi)
\left(\!\frac{\delta\phi_2}{\varphi_0}\!\right)^{\!\!2}\!\!-
9\xi\frac{\delta\phi_2}{\varphi_0} h_2 {+}
\frac{3y}{16x\xi}\!\left[10{-}\frac{3x\xi^2}{y}\right]\!h_2^2\right]\!.\!
\eeq
Both eigenvalues will be positive provided $\xi>0$ and 
\beq\label{eq:stab20}
\frac{y}{x}>\frac{3 \xi^2 (1 + 6 \xi)}{2 (5 + 12 \xi)}.
\eeq
Since $y>0$ is required for convergence of the EPI, this inequality
will be satisfied sufficiently near to the UVFP, i.e., to first order
in $\xi, y.$ All $n>2$ eigenvalues are also positive. There is no
guarantee that this continues to hold when nonlinear effects become
important, e.g., if the UVFP were approached along a trajectory in
violation of \eqn{eq:stab20}.

In sum, there are no unstable modes associated with the 
fluctuations, provided these inequalities are satisfied, as they
are near the UVFP. At one loop order, there are 6 zero modes (or flat 
directions.) One is the scalar dilaton, which we shall show gets mass at two 
loops. The other five are associated with a coherent fluctuation that, we 
believe, should be regarded as unphysical. 

Since we found no unstable modes, there will be no imaginary part to the 
one-loop correction. The result for the renormalized effective action is therefore 
the one given in our earlier paper~\cite{Einhorn:2015lzy} for this model, 
obtained by the renormalization group method.

The preceding remarks do not imply that the one-loop correction to the 
effective action cannot be negative at lower scales. In fact, our 
investigation~\cite{Einhorn:2015lzy} of the possibility of DT showed 
that it can indeed become negative. Unfortunately, we found the range
of couplings for which the extremum was actually a minimum did not lie
within the basin of attraction of the UVFP, so this model does not
produce a physically useful result. This was one of several reasons
that we proceeded to consider non-Abelian gauge theories, which are
potentially more physically relevant anyway.

\section{Transformation from the Jordan to the Einstein Frame}\label{sec:EFreal}

Most discussions of classical General Relativity proceed from the 
E-H action with minimal coupling. An action with non-minimal 
coupling, like the one discussed in the previous section, may under certain 
circumstances be transformed into a minimal coupling form by means of a 
conformal transformation of the metric. This is often referred to 
as going from the Jordan frame to the Einstein frame. Since this only involves 
a field redefinition, one might think that it is a simply matter of convenience, 
since interpretations of observables generally start from the Einstein frame. In 
the present context at least, we wish to argue that such a transformation is 
NOT so straightforward.

Given the classical action, \eqn{clJaction}, the conformal transformation is 
\beq\label{eq:conftransf}
\widetilde{g}_\mn\equiv\Omega^{-2} g_\mn,~~\mathrm{where}~~\Omega^2
\equiv \phi^2/{M^2},
\eeq
and $M$ is any convenient choice for the unit of mass. In any theory (and in 
the real world), the only observables are dimensionless ratios, so the choice for 
$M$ is arbitrary\footnote{In theories having other mass parameters, such as the 
Planck mass $M_P$ or scalar masses, $M$ is usually chosen to some 
combination of those parameters. We leave it unspecified for now.} but fixed 
(i.e., not scale dependent). Such a transformation is permissible provided 
$\Omega$ neither vanishes nor is singular. In classically scale-invariant 
models, this is not at all trivial. In the path integral, the integration over $\phi(x)$ 
is formally over all real values at every point, so it is impossible to guarantee 
this in general unless one assumes that it is a set of measure zero. This can be 
argued in the context of the perturbation expansion in which 
$\phi(x)=\vev{\varphi(x)}{+}\delta\phi(x),$ assuming that the background field 
$\vev{\varphi(x)}$ is nowhere vanishing and that the perturbation 
$\delta\phi(x)/\vev{\varphi(x)}$ is in some sense small, so that it makes sense
to assume $\varphi(x){\neq}0$ everywhere. Should the result of the 
calculation be that the on-shell background field vanishes anywhere, this 
construction would have to be revisited.
 
Assuming that $\phi(x)\ne0$, the effects of the field redefinition in 
\eqn{eq:conftransf} on the various quantities in \eqn{clJaction} are complicated. 
In Appendix~\ref{sec:conftransf}, we summarize the resulting changes on the 
various quantities entering the action. Defining $\xip\!\equiv\xi{+}1/6,$
$\zeta\!\equiv\!\sqrt{6\xip}M\log(\phi/M),$ and 
$\vartheta_\mu\! \equiv \pa_\mu\log\phi=
1/(\sqrt{6\xip}M)\pa_\mu\zeta,$ we find\footnote{We have dropped a 
surface term associated with $\nabla^2\zeta.$} 
\begin{subequations}
\label{eq:eaction3}
\begin{align} 
&S^{(E)}{=}\!\int\! d^4x \sqrt{{\wt{g}}} \bigg[\frac{\lambda M^4}{4}{-}
\frac{\xi M^2}{2}\wt{R} {+} \half\big(\wt{\nabla}\zeta\big)^{\!2} 
{+}\lcal_{ho}\bigg],\label{eq:emaction} \! \\
&\mathrm{where\ }
\lcal_{ho}\! \equiv \frac{1}{2a} \wt{C}^2{+}\frac{1}{3b} 
\big(\wt{R}{+}6\wt{\nabla}\!\cdot\!\vartheta 
{-}6\vartheta_\mu^2\big)^{\!2}
\!+c\,\wt{G}.
\label{eq:hoeinstein2}
\end{align}
\end{subequations}

On the one hand, we have simply performed a field redefinition, so one might 
expect the physics to be unchanged. On the other hand, the supposition that 
$\phi(x) \ne 0$ corresponds to SSB of scale 
invariance, so in fact, the physics is manifested quite differently in this broken 
phase. First of all, $\zeta$ plays the role of the dilaton, the 
(classical) Goldstone boson, which we previously identified in the Jordan frame 
from the mode expansion. (See discussion below \eqn{eq:mode00}.) As 
expected, $\zeta$ is derivatively coupled classically, so $\vev{\zeta},$
if constant, is arbitrary. (We shall find a convenient choice below in 
\eqn{vevzeta}.) In principle, in the QFT, it may or may not be the case that 
$\vev{\phi(x)} \neq 0.$ In fact, the issue of spontaneous breaking of scale 
invariance is actually moot in the QFT, because this is an anomalous 
symmetry. As a result, as mentioned earlier, this scalar will get a mass at 
two loops owing to the conformal anomaly.

The appearance of the dilaton field is just one consequence of the 
supposition that $\phi(x) \ne 0.$ The matter action, \eqn{eq:emaction}, 
takes the form of an E-H term linear in $\wt{R},$ with 
Planck mass-squared $M_P^2\equiv \xi M^2,$ plus a cosmological 
constant term with $M_P^2\Lambda\equiv \lambda M^4/4,$ plus a term 
corresponding to the kinetic energy of the dilaton $\zeta.$ 

The gravitational action, \eqn{eq:hoeinstein2}, involves, in addition to the 
quadratic curvature terms, involves terms in various powers of $\nabla\zeta/M.$
This is clearly extremely complicated, but it proves convenient to choose $M$ to 
be on the order of the SSB scale $v,$ where the one-loop correction has its 
minimum determined by DT. (See \reference{Einhorn:2015lzy}.) So long as 
$\xi(v)$ is in the range $0.1{-}10,$ this is also on order of the Planck mass, 
$M_P=\sqrt{\xi}M.$ For small dilaton momenta, more precisely, when 
$\sqrt{\xi}\,\wt{\Box}\zeta\ll \sqrt{1{+}6\xi} M_P \wt{R},$ these terms may be 
neglected in first approximation. Then the entire dependence on the dilaton 
field is given by the matter action, \eqn{eq:emaction}.

Although we have shown that DT can occur in this model, the values of the
coupling constants required for this to occur with local stability of the associated 
scale does not lie within the basin of attraction of the UVFP in this 
model~\cite{Einhorn:2015lzy}. Consequently, we shall defer the 
determination by DT and the calculation of the dilaton mass to the 
$SO(10)$ model in the next section.

\section{Non-Abelian Gauge Field}\label{sec:nonabelian}

We want to transform our $SO(10)$ model with a single adjoint 
scalar $\Phi$~\cite{Einhorn:2016mws} from the Jordan frame to the Einstein frame. 
Renormalizability will not be affected by a field redefinition and, for present 
purposes, a nonlinear transformation is useful. 
To review, our Jordan frame matter action is 
\begin{align}\label{eq:jmat}
\begin{split}
S^{(J)}_m=\int d^4x \sqrt{g}&\left[\frac{1}{4}\Tr[F_{\mu\nu}^2]+ 
\half\Tr[(D_\mu\Phi)^2]\right.\\
&\left. \hskip2mm -\frac{\xi \Tr [\Phi^2]}{2} R +V_J(\Phi)\right],
\end{split}
\end{align}
The adjoint scalar field $\Phi$ is a $10{\times}10$ Hermitian matrix that may be 
decomposed as $\Phi=\sqrt2\,\phi_a R^a,$ where the $\{\phi_a\}$ are real, and 
$\{R^a\}$ represents the 45 Hermitian generators of the fundamental or 
defining representation $\bf{10}$ of $SO(10).$ Similarly, the real, adjoint gauge 
field can be represented by $A_\mu=\sqrt2 A_\mu^aR^a,$ with the associated 
field strength 
$F_\mn\equiv\pa_\mu A_\nu{-}\pa_\nu A_\mu{-}i g[A_\mu,A_\nu]/\sqrt2.$ 
The covariant derivative of $\Phi$ is 
$D_\mu\Phi \equiv \pa_\mu\Phi{-}i g[A_\mu,\Phi]/\sqrt2.$ A brief review of our 
algebraic conventions is given in Appendix~\ref{sec:algebra}.

In order to transform to the Einstein frame, we want to presume that the model undergoes SSB $\vev{\Phi}\neq0.$ The exact nature of the 
breaking will be worked out in subsequent sections. A nonlinear field 
redefinition will enable us to proceed in much the same way as in the case of 
the real singlet in the preceding section. We define 
$T_2\equiv \Tr[\Phi^2]=\sum_a \phi_a^2,$ and define 
\beq\label{eq:omegasigma}
\Phi\equiv\Omega\,\vsig\ {\rm with}\ \Omega^2\equiv T_2/M^2,
\eeq 
in terms of an arbitrary unit of mass $M.$ Then,
\beq\label{vsigconstraint}
\Tr[\vsig^2]=M^2.
\eeq
Note that both $T_2$ and $\Omega$ are formally $SO(10)$ invariant.
 
One consequence of these definitions is that 
$\vev{\Phi}=\vev{\Omega}\vev{\vsig},$ so that $\vev{\Phi}\neq0$ if and
only if both $\vev{\Omega}\neq0$ and $\vev{\vsig}\neq0.$ Although one
may entertain other possibilities for SSB, they do not seem to be
relevant in perturbation theory. Then
\begin{subequations}\label{pavsig}
\begin{align}
D_\mu\Phi&=\vsig\,\pa_\mu\Omega+\Omega\,D_\mu\vsig\\
\Tr[(D_\mu\Phi)^2]&=
M^2(\pa_\mu\Omega)^2+\Omega^2\Tr[(D_\mu\vsig)^2].
\end{align}
\end{subequations}
In passing from the first to the second line in \eqn{pavsig}, the cross term 
vanishes because 
\begin{align}\label{dconstraint}
\begin{split}
\Tr[\vsig D_\mu\vsig]=
\Tr[\vsig\pa_\mu\vsig]
{-}ig\Tr\big[\vsig[A_\mu,\vsig]\big]&=\\
\pa_\mu\Tr[\vsig^2]/2=\pa_\mu M^2/2&=0,
\end{split}
\end{align}
where the term involving the gauge field $A_\mu$ vanishes by the cyclic 
property of the trace. The Jordan frame Lagrangian density, \eqn{eq:jmat}, then 
becomes
\begin{align}\label{eq:jmat2}
\begin{split}
\lcal^{(J)}_m&=\sqrt{g}\bigg[\frac{1}{4}\Tr[F_\mn^2]-\frac{\xi M^2\Omega^2}{2} R+\\
&\hskip5mm \frac{M^2}{2}(\pa_\mu\Omega)^2{+}
\frac{\Omega^2}{2}\Tr[(D_\mu\vsig)^2{+}V_J(\Omega\,\vsig)]\bigg]\!,\!
\end{split}
\end{align}
subject to the constraint $\Tr[\vsig^2]=M^2,$ \eqn{vsigconstraint}.

The original field $\Phi$ provided a linear representation of a real
adjoint multiplet and represented 45 DoF in the matter action
\eqn{eq:jmat}. Evidently, in \eqn{eq:jmat2}, one degree of freedom has
been apportioned to $\Omega$ and only 44 DoF remain in $\vsig.$ This
can be seen from \eqn{dconstraint}, which implied that
$\Tr[\vsig\pa_\mu\vsig]=0.$ Thus, the dynamical degrees of freedom
associated with $\pa_\mu\vsig$ are restricted to those ``orthogonal" to
$\vsig.$ 

To complete this rewriting of the action \eqn{eq:jmat}, consider the potential, 
$V_J(\Phi).$ Defining 
$T_4\equiv \Tr[\Phi^4]=\Omega^4\Tr[\vsig^4],$ the potential is 
\begin{align}
V_J (\Phi)\!\equiv \frac{h_1}{24} T_2^{\,2}{+}\frac{h_2}{96}T_4= 
\frac{h_1M^4}{24} \Omega^4{+}\frac{h_2}{96} \Omega^4\Tr[\vsig^4].\!
\end{align}
Thus, the only dependence of $V_J$ on $\vsig$ is through $T_4.$

Further, the nonminimal coupling to the curvature in \eqn{eq:jmat2} is 
independent of $\vsig.$ As a result, the $SO(10)$ singlet $\Omega$ plays the 
role of the real scalar in the preceding section. Evidently, to transform to 
the Einstein frame, we need only suppose that $\vev{\Omega}\neq0$ and can 
postpone the question of $\vev{\vsig}$ until later. Without loss of generality 
(WLOG), we take $\vev{\Omega}>0.$ Then we can perform a conformal 
transformation, $\wt{g}_\mn\equiv \Omega^{-2} g_\mn,$ to get the action in 
the Einstein frame
\begin{subequations}\label{einsteinframe}
\begin{align}\label{eq:esho}
S^{(E)}_{ho}&{=}\!\!\int\!\! d^4x\sqrt{\wt{g}} \Bigg[\frac{\wt{C}^2}{2a}{+}
\frac{1}{3b}\big(\wt{R}{+}6\wt{\nabla}\!\cdot\!\vartheta 
{-}6\vartheta_\mu^2\big)^{\!2}{+}c\,\wt{G}\Bigg]\!,\!\\
\begin{split}\label{eq:emat}
S^{(E)}_m&=\!\!\int\!\! d^4x\sqrt{\wt{g}}\left[\frac{1}{4}\Tr[\wt{F}_\mn^2]
-\frac{\xi}{2} M^2 \wt{R}+\frac{(\wt{\nabla}\zeta)^2}{2}+
\right. \\
&\left. \hskip8mm \frac{h_1}{24}M^4+
\half\Tr[(\wt{D}_\mu\vsig)^2]+\frac{h_2}{96}\Tr[\vsig^4]\right],\!
\end{split}
\end{align}
\end{subequations}
where, similar to the previous case, \eqn{eq:eaction3}, 
\beq\label{eq:zeta}
\zeta \equiv M\sqrt{6\xip}\log\Omega,\ \  \vartheta_\mu \equiv
\wt{\pa}_\mu\log{\Omega}=\frac{1}{\sqrt{6\xip}M}\wt{\pa}_\mu\zeta.
\eeq
We must also keep in mind the constraint, \eqn{vsigconstraint}. 

One can show that the G-B term changes by a total divergence, 
$G{\to}\wt{G}{+}\nabla_\mu J^\mu.$ (See Appendix~\ref{sec:conftransf}.)
Although not the simplest form to quantize, the spectrum may be read off
rather easily. The last line of \eqn{eq:emat} shows that $\vsig$ is
described by a gauged nonlinear sigma model with a scale-invariant
self-interaction strength proportional to $h_2.$ Regardless of the pattern of 
SSB, $\Tr[\vsig^4]\geq (\Tr[\vsig^2])^2/10=M^4/10,$ and it proves useful 
to rewrite the terms involving $h_1, h_2$ as the sum of two nonnegative terms
\beq\label{eq:mattpot}
\frac{h_3M^4}{24}+\frac{h_2}{96}\left[\Tr[\vsig^4]{-}\frac{M^4}{10}\right],
\eeq
where we defined $h_3\equiv h_1{+}h_2/40.$

Note that the action $S^{(E)}=S^{(E)}_{ho}{+}S^{(E)}_m$ is still formally 
invariant under 
the $SO(10)$ local gauge symmetry, since the conformal transformation 
employed only the gauge singlet $\Omega(x),$ which was presumed to have 
some non-zero vacuum expectation value (VEV) $\vev{\Omega(x)}$ to be 
determined. Although in principle, this can vary with position $x^\mu,$ there is 
a tacit assumption that $\Omega(x)$ vanishes nowhere since, otherwise, the 
transformed metric would degenerate somewhere. For simplicity, we shall seek 
SSB solutions in which $\vev{\Omega}\neq0$, independent of $x.$

The role of the couplings $h_1, h_2,$ (or $h_2, h_3$) in the Einstein frame 
suggests a dramatically different physical picture than that in the Jordan frame. 
In \eqn{eq:emat}, we can identify the Planck mass
\beq\label{mplanck}
M_P=\sqrt{\xi}\,M.
\eeq
As in the case of the real field, we must have $\xi(v)>0$ at the scale $v$ of 
symmetry breaking in order for gravity to be attractive.

From \eqn{eq:mattpot}, the ``vacuum energy density" is $h_3 M^4/24$ or 
possibly larger, depending on the direction of SSB $\vev{\vsig}.$ In more 
conventional terms, the cosmological constant $\Lambda$ corresponding to a 
vacuum energy density equal to $h_3M^4/24$ is 
\beq\label{eq:cc}
\Lambda\equiv \frac{h_3}{24\xi^2} M_P^2.
\eeq
Thus, we must have $h_3(v)>0$ at the scale of symmetry-breaking in order for 
$\Lambda$ to be positive. 

The field $\zeta$ is the dilaton, which is massless in this approximation but will 
 gain mass at two-loop order, $O(\hbar^2).$ (We shall determine its 
mass below in \secn{sec:dilaton}.) 

This is as much as can be said at the classical level about the singlets in 
\eqn{einsteinframe}. Further interpretation requires knowing more precisely the 
pattern of the breaking of $SO(10),$ which we shall discuss next.

\section{Spontaneous Symmetry Breaking of $\bf{SO(10)}$}
\label{sec:ssb}

In our parameterization, the direction of $SO(10)$ breaking is embodied in 
$\vev{\vsig}.$ Since the $\vsig$ field enters the Einstein frame action 
only via \eqn{eq:emat}, we can determine the possible extrema ignoring 
\eqn{eq:esho}, which is to say that they are essentially independent of the scale 
of SSB. In fact, we already showed in \reference{Einhorn:2016mws} that the 
only extremum that is a local minimum corresponds to breaking to $\su51.$ In 
passing to the Einstein frame, we only utilized the singlet field $\Omega(x),$ so we 
would not expect this pattern to change. Indeed, unless there exists a sensible 
phase in which $\vev{\Phi}=0,$ the Einstein frame action, \eqn{einsteinframe}, 
must be completely equivalent to the Jordan frame action, 
$S^{(J)}\equiv S_{ho}^{(J)}+S^{(J)}_m,$ \eqns{eq:Jsho}{eq:jmat}. 
Although we could proceed by assuming this pattern of SSB is correct, it is 
illuminating to rederive it in the Einstein frame to confirm this expectation and to 
take note of the substantial differences from the Jordan frame.

For our purposes, it is convenient to make a unitary transformation to a basis 
in which the generators of $SO(10)$ take the form
\beq\label{eq:basis}
R^a\equiv
\frac{1}{\sqrt2}\!
\begin{pmatrix}
\ \rcal^a_1 &\vline& \rcal^a_2 \\
\hline
 \rcal^a_2{}^\dag &\vline&\!\! -\rcal^a_1{}^\tau
\end{pmatrix}
{=}\frac{1}{\sqrt2}\!
\begin{pmatrix}
\rcal^a_1 &\vline& \rcal^a_2 \\
\hline
-\rcal^a_2{}^*\! &\vline&\!\! -\rcal^a_1{}^*
\end{pmatrix}\!,\!
\eeq
where\footnote{The factor $1/\sqrt{2\!}$ has been inserted so that $\!\{\rcal_1^a\}\!$ 
are the generators of $\su51$ with canonical normalization, 
$\Tr[\rcal_1^a\rcal_1^b]=\delta_{ab}/2$ for $a{=}\{1,2,\ldots,25\}.$ For further 
discussion, see Appendix~\ref{sec:algebra}.} the $ \rcal^a_j$ are $5{\times}5$ 
(complex) matrices with the properties that $\rcal_1^a$ is Hermitian, and 
$\rcal_2^a$ is antisymmetric. Hence, $\rcal_1^a{}^*=\rcal_1^a{}^\tau,$ where 
$\tau$ denotes the transpose. In this basis, unlike the original one, the Cartan 
subalgebra of $SO(10)$ can be diagonalized.

Correspondingly, we define for real components $\sigma_a$
\begin{subequations} \label{so10sigma}
\begin{align}
\vsig\equiv \sqrt2\sigma_a R^a
&=\begin{pmatrix}
\sigma_a \rcal^a_1\! &\vline&\!\! \sigma_a \rcal^a_2 \\
\hline
 \sigma_a\rcal^a_2{}^\dag \!&\vline&\!\! - \sigma_a\rcal^a_1{}^*
\end{pmatrix}\\
&\label{eq:vsig} \equiv 
\begin{pmatrix}
\vsig_1 &\vline&\!\!\vsig_2 \\
\hline
 \vsig_2{}^\dag &\vline&\!\! {-}\vsig_1{}^*
\end{pmatrix}.
\end{align}
\end{subequations} 
The constraint \eqn{vsigconstraint} implies 
\beq\label{vsigmaconstraint}
\sum_1^{45} \sigma_a^2=M^2,\ \ \mathrm{or}\ 
\Tr[\vsig_1^2+\vsig_2^\dag\vsig_2]=M^2/2.
\eeq
Assuming $\vev{\vsig}\ne0,$ one may utilize the $SO(10)$
symmetry of the action \eqn{einsteinframe} to bring it to diagonal form
\beq\label{eq:vevSigma}
\vev{\vsig}=
\begin{pmatrix}
\vev{\vsig_1} & \vline&\, 0 \\
\hline
0 \! & \vline&\! \!{-}\vev{\vsig_1}
\end{pmatrix},
\eeq
where $\vev{\vsig_1}$ is the matrix, ${\mathrm{Diag}}
\{\varsigma_1,\varsigma_2,\varsigma_3,\varsigma_4,\varsigma_5\}.$
These $\varsigma_i$ are the eigenvalues of $\vev{\vsig},$ which are of course 
independent of the choice of basis. However, the basis chosen above, 
\eqn{eq:basis}, is particularly convenient. Let us call the generators of the 
$SO(10)$ Cartan subalgebra $H^i,$ with the corresponding generators of 
$\su51$ $\hcal^i.$ Then we conclude that
\beq\label{eq:vevCartan}
\varsigma_iH^i=\vev{\sigma_a}R^a,\ {\rm and}\ \ 
\varsigma_i\hcal_1^i=\vev{\sigma_a}\rcal_1^a.
\eeq

With reference to the action \eqn{eq:emat} and the constraint
\eqn{vsig1constraint}, in order to seek the extrema of the action, we must 
consider 
 \beq\label{eq:lagrange}
 \frac{h_2}{48}\Tr[\,\vev{\vsig_1}^{\!4}]{-}\frac{\eta}{2} \Tr[\,\vev{\vsig_1}^{\!2}]=
 \sum_{i=1}^5\bigg(\frac{h_2}{48}\varsigma_i^4{-}
 \frac{\eta}{2}\varsigma_i^2\bigg),
 \eeq
 where $\eta$ is a Lagrange multiplier associated with the constraint
 \beq\label{vsig1constraint}
 \Tr[\, \vev{\vsig_1}^2]=M^2/2.
 \eeq
 The first derivative of \eqn{eq:lagrange} is
\beq\label{eq:first}
\frac{h_2}{12}\varsigma_i^3-\eta\varsigma_i=
\varsigma_i\left(\frac{h_2}{12}\varsigma_i^2-\eta\right),\ \ \{i=1,\ldots, 5\}.
\eeq
This will vanish for each $\varsigma_i$ provided either $\varsigma_i=0$ or 
$\varsigma_i=\pm\varsigma_0,$ where $\varsigma_0\equiv\sqrt{12\eta/h_2}.$ With 
regard to the sign of the nonzero $\varsigma_i,$ it can be resolved as we did in 
the Jordan frame~\cite{Einhorn:2016mws}. Referring to \eqn{eq:vevSigma}, by 
means of a unitary transformation, we may interchange any negative element 
in $\vev{\vsig_1}$ with the corresponding positive element in $-\vev{\vsig_1}.$ 
Thus, WLOG, we may assume that the elements of $\vsig_1$ are nonnegative. 
There are then five distinct extrema, depending on the number $k$ of zeros in 
$\vsig_1,$ so that $T_2=2(5{-}k)\varsigma_0^2=M^2,$ $k{=}\{0,1,\ldots,4\}.$ 
Therefore\footnote{$\varsigma_0$ implicitly depends upon $k$, but we hope 
that will be clear in context without having to introduce more cumbersome 
notation.}, $\varsigma_0=\sqrt{1/(2(5{-}k))}M,$ with corresponding Lagrange 
multiplier $\eta=h_2M^2/(24(5{-}k)){>}0.$

To determine which of these five extrema are minima, we consider 
the second derivative is
\beq\label{eq:second}
\frac{\pa^2V(\sigma)}{\pa\varsigma_i\pa\varsigma_j}=
\delta_{ij}\left[\frac{h_2}{4}\varsigma_i^2-\eta\right].
\eeq
This is diagonal as well (unlike the Jordan frame 
calculation~\cite{Einhorn:2016mws}) with elements either $-\eta$ if 
$\varsigma_i=0$ or $h_2 \varsigma_0^2/4-\eta$ if $\varsigma_i\ne0.$ 
Since $\eta>0,$ the extremum has an unstable mode if \emph{any} of the 
$\varsigma_i$ is zero. Taking $k=0$ then, we find that
$V''(\varsigma_0)=h_2\varsigma_0^2/4{-}\eta=h_2M^2/40-h_2M^2/120=
h_2M^2/60>0,$ so this case is (locally) stable, just as 
before~\cite{Einhorn:2016mws}. Hence, 
\beq\label{eq:vevvsig1}
\vev{\vsig_1}=\varsigma_0{\bf 1_5}, \mathrm{with}\ \varsigma_0=M/\sqrt{10}.
\eeq

In sum, we have confirmed that, in this model, the only possibility for SSB to a 
phase having a local minimum is $SO(10){\to}SU(5){\otimes} U(1).$ Having 
done so, we are now in a position to determine the masses of the vector bosons 
and the other heavy scalars arising from fluctuations in~$\varSigma.$

\section{Vector Boson Masses}\label{sec:vbmass}

Another quantity that can be read directly from the Einstein frame action 
\eqn{einsteinframe} is the mass of the vector bosons, which, in this section, 
will be shown to be $M_V=g M_P/\sqrt{5\xi}.$ 
These masses arise from the scalars' covariant derivative in \eqn{eq:emat}
\beq\label{eq:Dvsig2}
\half\Tr \left[D_{\mu}\vsig\right]^2 = 
\half{\sum_a} \left(\pa_{\mu}\sigma_a +gf^{abc}A_b^\mu\sigma_c\right)^2,
\eeq
using $\vsig = \sqrt{2}\sigma_aR^a$, as in \secn{sec:nonabelian}. The $f^{abc}$ 
are the structure constants for $SO(10).$ 
In the action, \eqn{eq:emat}, the field strength, $\Tr[F_\mn^2]/4,$ is canonically 
normalized. Therefore, \eqn{eq:Dvsig2} implies that the vector boson mass matrix is 
\beq\label{eq:vbmatrix}
(M_V^2)_{ab} = g^2 f^{acd}f^{bce}\vev{\sigma_d}\vev{\sigma_e},
\eeq
As discussed in Appendix~\ref{sec:algebra}, the 20 gauge bosons that acquire 
mass after SSB transform as conjugates 
$\bf{10_4}{\oplus}\bf{\overline{10}_{-4}}$ of $\su51.$ As a result, all 20 will have 
the same mass $M_V,$ so we may simplify the calculation by summing 
\beq
\Tr \big[\vev{(M_V^2)}\big] = g^2C_G\textstyle{\sum_a}\vev{\sigma_a}^2=
g^2 C_G\Tr[\,\vev{\vsig}^2],
\eeq
where $C_G$ is the quadratic Casimir in the adjoint representation.
In $SO(N)$, $C_G = (N-2)/2,$ so 
\beq
\Tr [(M_V^2)] = 4g^2\Tr[\,\vev{\vsig}^2]=4g^2M^2=4g^2M_P^2/\xi,
\eeq
where, in the last steps, we applied first the constraint \eqn{vsigconstraint} 
and then \eqn{mplanck}. As mentioned, the 20 particles have identical masses, 
so each of them has mass 
\beq\label{eq:vbmass}
M_V = gM_P/\sqrt{5\xi},
\eeq

\section{Heavy Scalar Masses}\label{sec:scalarmass}

Unlike the 25 massless vector bosons, the $\su51$ gauge 
symmetry does not protect the 25 corresponding adjoint scalars 
from acquiring invariant masses after SSB of $SO(10).$ Returning to the
action, \eqn{eq:emat}, we have previously mentioned that formally it
remains $SO(10)$ gauge invariant. As an aside, one might think it would
permit $\vev{\vsig}=0,$ but that is illusory, a result of using a
nonlinear representation of the symmetry. As discussed in 
\secn{sec:ssb}, the transformation to the Einstein frame tacitly requires 
$\vev{\vsig}\neq0,$ and that property is also subsumed in the constraint
conditions \eqns{vsigconstraint}{vsig1constraint}. Thus, despite
appearances, $SO(10)$ must be spontaneously broken to arrive at
\eqn{eq:emat}.

To determine the scalar masses, we shall start from the decomposition of 
$\vsig$ into block form, \eqn{eq:vsig}. It is convenient to work in a gauge (e.g.\  
unitary gauge) in which the off-diagonal blocks involving $\vsig_2$ have been 
``eaten" to give masses to the vector bosons, so that $\vsig_2=0.$ Then $\vsig$ 
takes the form: 
\beq\label{eq:blocksigma}
\vsig=\sqrt2\sigma_aR^a=
\begin{pmatrix}
\ \vsig_1&\vline&0 \\
\hline
0 &\vline&\!\! -\vsig_1^*
\!\end{pmatrix},
\eeq
with $\vsig_1=\sigma_a\rcal_1^a.$ ({\it N.B.} $\vsig_1$ is not diagonal.) 
As explained in \secn{sec:nonabelian}, $\vsig$ has only 44 independent DoF 
before SSB. With 20 absorbed by the vector bosons, only 24 DoF remain in 
$\vsig_1.$ 

This results in a fundamental difference\footnote{The reader may wish to refer 
to the branching rules for $SO(10){\to}SU(5){\otimes} U(1),$ 
\eqn{eq:branching}.} 
between the masses of the $\su51$ scalar multiplet ${\bf 24_0}$ associated 
with $\vsig$ or $\vsig_1$ and the singlet $\bf{1_0}$ attributed to $\Omega$ 
through $\zeta.$ 

We wish to solve the constraint conditions to make explicit the 24 DoF 
represented by $\vsig.$ Writing the adjoint field $\vsig$ in the form 
$\vsig=\vev{\vsig}{+}\vdelvsig,$ we may expand the Einstein frame action 
\eqn{einsteinframe} about the background $\vev{\vsig},$ assumed as usual to be 
constant. Keeping only the terms depending on $\vsig,$ the Lagrangian density becomes
\beq\label{eq:vsiglag}
\lcal_S \equiv \half\Tr\big[(D_\mu\vdelvsig)^2\big]+
\frac{h_2}{96}\Tr\Big[\big(\vev{\vsig}{+}\vdelvsig\big)^4\Big].
\eeq
To determine the masses associated with $\vdelvsig,$ we may 
neglect the gauge bosons in \eqn{eq:vsiglag} and expand the potential 
terms through quadratic order in $\vdelvsig.$ Recall from 
\secn{sec:left} that $\vev{\vsig_1}=\varsigma_0{\bf 1_5},$ with 
$\varsigma_0=M/\sqrt{10}.$ Then \eqn{eq:vsiglag} becomes
\begin{align}\label{eq:delvsiglag}
\begin{split}
&\hskip-5mm \lcal_S = \Tr\big[(\pa_\mu\vdelvsig_1)^2\big]{+}
\frac{h_2}{48}\Big[ 5\varsigma_0^4{+}\\
&\hskip15mm 
4\varsigma_0^3\Tr[\vdelvsig_1]{+} 
6\varsigma_0^2\Tr[\vdelvsig_1^2]{+}\ldots\Big].
\end{split}
\end{align}
The normalization of the kinetic energy in \eqn{eq:delvsiglag} appears to be 
not canonical, but, from \eqn{eq:blocksigma}, 
$\vdelvsig_1=\delta\sigma_a\rcal_1^a$ 
and $\Tr[\rcal_1^a\rcal_1^b]=\delta_{ab}/2.$ Therefore,
\beq
\Tr[(\pa_\mu\vdelvsig_1)^2]=
\half\!\sum_1^{24}(\pa_\mu\delta\sigma_a)^2.
\eeq

We must now take into account the constraints \eqns{vsigconstraint}{vsig1constraint},
\begin{subequations}\label{eq:constraint2}
\begin{align}
&\Tr[\big(\vev{\vsig_1}+\vdelvsig_1\big)^2]=M^2/2,\ {\rm or}\\
&2\varsigma_0\Tr[\vdelvsig_1]+\Tr[(\vdelvsig_1)^2]=0.
\end{align}
\end{subequations}
To interpret this constraint, we decompose the 25 components of 
$\vdelvsig_1$ as
\beq\label{eq:decompose}
\vdelvsig_1=\frac{\varDelta{S_1}}{5}{\bf 1_5}+
\varDelta\wt{\vsig}_1,
\eeq
with\footnote{Although $\varDelta{S_1}$ transforms as a $\su51$ singlet, it
must not be confused with the $\bf{1_0},$ $SO(10)$ singlet $\Omega$ 
(or $\zeta).$} $\varDelta S_1\!\equiv\!\Tr[\vdelvsig_1],$ so that 
$\Tr[\varDelta\wt{\vsig}_1]\!=\!0.$ Then \eqn{eq:constraint2} implies
\begin{subequations}\label{eq:dels1}
\begin{align}
&2\varsigma_0\varDelta{S_1}{+}\frac{1}{5}{\varDelta{S_1}}^2{+}
\Tr\big[(\varDelta\wt{\vsig}_1)^2\big]=0,\\
&{\implies}\frac{\varDelta{S_1}}{5\varsigma_0}=\sqrt{1{-}
2\Tr[(\varDelta\wt{\vsig}_1)^2]/M^2} -1 \approx \label{eq:sqrt}\\
&\hskip5mm {-}\Tr[(\varDelta\wt{\vsig}_1)^2]/M^2{+}
\Tr[(\varDelta\wt{\vsig}_1)^2]^2/(2M^4){+}\ldots.\label{eq:delvsig1}
\end{align}
\end{subequations}
In \eqn{eq:sqrt}, the positive square root must be chosen so that 
$\varDelta{S_1}{\to}0$ for $\varDelta\wt{\vsig}{\to}0.$ The interpretation of 
\eqn{eq:dels1} is that $\varDelta{S_1}$ is determined 
by $\varDelta\wt{\vsig}_1$ of $SU(5)$, with the leading term 
of $\varDelta{S_1}$ being quadratic in $\varDelta\wt{\vsig}_1.$

Returning to $\lcal_S$ in \eqn{eq:delvsiglag}, we want to decompose 
$\vdelvsig_1$ as in \eqn{eq:decompose} and replace $\varDelta{S_1}$ using
\eqn{eq:delvsig1}. First, in the kinetic term, $(\pa_\mu\varDelta{S_1})^2$ can be 
discarded since it is actually fourth order in $\varDelta\wt{\vsig}_1.$ 
Next, the second line of \eqn{eq:delvsiglag} can be reexpressed as
\beq
4\varsigma_0^3\varDelta{S_1}{+}6\varsigma_0^2
\Tr\big[(\varDelta\wt{\vsig}_1)^2\big]
\approx 4\varsigma_0^2\Tr\big[(\varDelta\wt{\vsig}_1)^2\big]{+}\ldots,
\eeq
neglecting terms in $\varDelta\wt{\vsig}_1$ of higher order than quadratic. 

Since the kinetic term is canonically normalized, the mass-squared of the 
24 $SU(5)$ adjoint scalars is
\beq
M_{\!\vdelvsig}^{\;2}=4\varsigma_0^2=\frac{2}{5}M^2= 
\frac{2}{5\xi}M_P^2=\frac{2}{g^2}M_V^2,
\eeq
where we have included the last two relations in order to facilitate 
comparison of these scalar masses with the Planck mass and the massive 
gauge bosons.

\section{The Background Curvature and its Fluctuations}\label{sec:rho}

Renormalizable gravity is a scalar-tensor theory of gravity, \ie the metric involves a 
scalar DoF in addition to the usual tensor degree of freedom associated with the 
graviton. The scalar DoF can be identified with fluctuations of the scalar curvature 
$\wt{R}.$ In the Jordan frame, we had to calculate the radiative corrections to the effective 
action to determine the magnitude of the background curvature $\vev{R_J}$ and its 
fluctuations. Therefore, it may come as a surprise that the first approximation to the 
corresponding quantity in the Einstein frame can be calculated from the 
classical action \eqn{einsteinframe}. On second thought, since \eqn{eq:emat} 
contains a cosmological constant, both before and after SSB, one could have 
anticipated that $\wt{R}=4\Lambda$ already from the matter action. Therefore, 
before getting into the matter of calculating radiative corrections to the effective 
action, we shall first discuss the tree approximation. 

For this purpose, as well as to enable calculation of radiative corrections in the next 
section, we must make some simplifying assumptions. By defining
$\Phi\equiv\Omega\vsig,$ we have distinguished the magnitude $\Omega$ of 
$\Phi$ from its direction $\vsig.$ At the classical level, \eqn{einsteinframe}, 
we saw that the two fields were essentially decoupled, \ie $\Omega$ is 
expressed through the dilaton field which does not couple directly to $\vsig.$ 
Consequently, we may replace $\vsig{\to}\vev{\vsig},$ while still keeping the 
background metric and $\zeta$ (or $\Omega$) off-shell. In general, the metric 
$\wt{g}_\mn$ couples directly to everything via the factor $\sqrt{\wt{g}}$ in the 
invariant volume density, but this will not cause problems in leading order. 
$\vsig$ also couples to the metric through the kinetic term 
$(\pa_\mu\vsig)^2$ but, assuming that $\vev{\vsig}$ is constant, 
$\pa_\mu\vev{\vsig}=0.$ 

Secondly, we shall assume that the background metric has maximal global 
symmetry, \viz that of dS spacetime. In that case, the Euclidean spacetime 
volume is 
\beq\label{volume}
\int d^4x\sqrt{\vev{\wt{g}}}=\bigg(\frac{12}{\wt{R}}\bigg)^{\!\!2}\, 
\frac{8\pi^2}{3}\equiv \frac{V_4}{\rho^4},
\eeq
where, for economy of writing henceforth, we defined 
$\rho\equiv(\wt{R})^{1/2}$ 
and the unit volume $V_4\equiv 12^2 {\times} 8\pi^2/3=384\pi^2.$ This does 
not require $\rho$ to be on-shell; it follows simply from the assumption that the 
background has maximal symmetry, so that the spacetime volume is a sphere 
$S^4,$ with an arbitrary radius of curvature related to $\rho.$ 

Our goal is to determine the extrema of $\rho$ and $\zeta$ and to determine 
which are minima. Our first task is to determine that they have stable, constant 
background fields. So for the moment, we shall assume that both are constant. 
With these assumptions, the classical action \eqn{einsteinframe} is 
independent of $\zeta,$ since it only has derivative couplings. As we have 
explained in \secn{sec:ssb}, this is because classically, $\zeta$ is a Goldstone 
boson. It is only from quantum corrections that we can determine whether 
$\zeta$ has a minimum, even for constant $\zeta.$ However, unlike the Jordan 
frame calculation that appears to only enable one to determine the ratio of 
fields\footnote{In fact, because of the mixing between modes, we 
discovered in \reference{Einhorn:2016mws} only belatedly that the minimum in 
$\rho$ called $\varepsilon_1$ was classical. 
The calculation in this section makes 
that clear from the outset.}, we can, 
as a consequence of the conformal transformation,
determine the minimum in $\rho$ directly 
from the classical action. 
Under the preceding assumptions, the value of the classical action 
\eqn{einsteinframe} off-shell is 
\beq\label{eq:vevsecl}
\frac{1}{V_4}S^{(E)}_{cl}\!(\rho)=\left[\frac{1}{3b}+
\frac{c}{6}-\frac{\xi M^2}{2\rho^2} 
+\frac{h_3M^4}{24\rho^4}\right].
\eeq

It may be surprising at first sight that the contributions from the higher-
order action, \eqn{eq:esho}, are independent of $\rho.$ It is clear that setting 
$\vsig{\to}\vev{\vsig}$ affects only \eqn{eq:emat}, but the curvature and dilaton 
fields enter \eqn{eq:esho} as well. Upon reflection, this observation results 
from the assumption that the classical background fields have constant 
curvature $\rho$ (or $\wt{R}$) and constant $\zeta$ (or $\Omega$). Then the 
higher-derivative action, \eqn{eq:esho}, has the same classically scale 
invariant form as in the Jordan frame. Even off-shell, its value is independent of 
$M,$ and, being dimensionless, it must also be independent of $\rho.$

To determine the extrema in $\rho$ and its nature, we calculate the first two derivatives of \eqn{eq:vevsecl}:
\begin{subequations}\label{vary12}
\begin{align}
\label{eq:firstcl}
\frac{1}{V_4}\frac{\pa S_{cl}^{(E)}}{\pa \rho}
&=\frac{\xi M^2}{\rho^3}-\frac{h_3 M^4}{6\rho^5},\\
\label{eq:secondcl}
\frac{1}{V_4}\frac{\pa^2 S_{cl}^{(E)}}{\pa \rho^2}&=-\frac{3\xi M^2}{\rho^4}
+\frac{5 h_3 M^4}{6\rho^6}.
\end{align}
\end{subequations}
\eqn{eq:firstcl} vanishes for 
\beq\label{eq:rho0}
\rho_0^2= \frac{h_3}{6\xi} M^2=\frac{h_3}{6\xi^2}M_P^2. 
\eeq
For $\rho=\rho_0,$ the curvature \eqn{eq:secondcl} of the potential 
becomes $2\xi M^2/\rho_0^4=2M_P^2/\rho_0^4{>}0,$ so $\rho_0$ is in 
fact a minimum of the classical potential. To translate this into a mass 
parameter, we return to the Lagrangian density by dividing the action by 
the invariant spacetime volume \eqn{volume}.  Expanding 
$\rho=\rho_0{+}\delta\rho$ to second order in $\delta\rho,$
\begin{subequations}\label{rhomass}
\begin{align}
S_{cl}^{(\!E)}\!(\rho)&=S_{cl}^{(\!E)}\!(\rho_0)+
\!\int\! d^4x\sqrt{\wt{g}}\ \frac{m_\rho^2}{2}\,\delta\rho^2+\ldots,\\
\mathrm{with}\ m_\rho&\equiv\sqrt2 M_P.
\end{align}
\end{subequations}

For future reference, the on-shell value of the classical action in the Einstein 
frame, \eqn{eq:vevsecl}, is
\beq\label{onshellscl}
\frac{1}{V_4}S_{cl}^{(\!E)}\!(\rho_0)=\left[\frac{1}{3b}{+}\frac{c}{6}{-}
\frac{\xi M^2}{4\rho_0^2}\right]{=}\left[\frac{1}{3b}{+}\frac{c}{6}{-}
\frac{3\xi^2}{2h_3}\right].
\eeq

\eqn{rhomass} is only valid for constant fluctuations $\delta\rho;$ however, 
we need to demonstrate stability for non-static fluctuations of the background. 
This becomes complicated, even assuming that the background is dS 
spacetime with constant curvature $\wh{R}{=}\rho_0^2.$ So long as we 
assume that the background $\zeta$ is constant, this analysis can be carried 
out classically by expanding \eqn{einsteinframe} to second-order in metric 
fluctuations. To express local fluctuations in the metric, we follow the same 
path as in \secn{sec:jordanreal}, writing 
$\wt{g}_\mn \equiv \! \wh{g}_\mn{+}h_\mn,$ with the background $\wh{g}_\mn$ 
describing dS spacetime with constant curvature $\wh{R}=\rho_0^2,$ and 
$h_\mn(x)$ corresponding to the fluctuations. For $h_\mn(x),$ we adopt the 
transverse-traceless (TT) gauge, described in \eqn{eq:TT} {\it et seq.} 
To explore stability, we need to expand the fluctuations through second 
order, up to which there is no mixing between the fluctuations of fields 
having nontrivial classical backgrounds and those that do not.  The 
dilaton field is exceptional, inasmuch as it only appears in 
\eqn{einsteinframe} derivatively coupled. In that case, its fluctuations still do not mix with other fields to quadratic order. 
Assuming that the background 
vector field $A_\mu$ vanishes, the result for fluctuations to the metric 
will, to quadratic order, be the same \emph{as if} we started from the 
classical action
\begin{subequations}\label{einsteinframe2}
\begin{align}\label{eq:esho2}
S^{(E)}_{ho}&{=}\!\!\int\!\! d^4x\sqrt{\wt{g}} \Bigg[\frac{1}{2a}\wt{C}^2{+}
\frac{1}{3b}\wt{R}{}^2{+}c\,\wt{G}\Bigg]\!,\!\\
\label{eq:emat2}
S^{(E)}_m&=\!\!\int\!\! d^4x\sqrt{\wt{g}}\left[
-\frac{\xi M^2}{2}  \wt{R}+
\frac{h_3}{24}M^4\right],
\end{align}
\end{subequations}
This is precisely the action for renormalizable gravity with the inclusion 
an explicit Planck mass, \eqn{mplanck} and a cosmological constant, 
\eqn{eq:cc}. This may be obtained from the model discussed in 
\secn{sec:jordanreal} for the real field, \eqn{clJaction} with the replacements 
\beq\label{substitute1}
\delta\phi\to0, \quad \varphi \to M, \quad \lambda \to h_3/6.
\eeq
This model was previously analyzed by Avramidi~\cite{Avramidi:1986mj}. 
As we mentioned in the Introduction, \secn{sec:intro}, by expanding in the 
Jordan frame, he showed that, with the exception of the five zero modes 
that we discussed earlier, the fluctuations are stable for a certain range 
of coupling constants, a result that seems not to be as well known as 
perhaps it should be. This conclusion should apply to the Einstein frame 
on-shell, since the difference in the actions between the two frames is 
simply a field redefinition\footnote{For further discussion, see \eg  
\reference{Flanagan:2004bz}.}. Therefore, we can simply adapt Avramidi's 
results\footnote{See Eqs. (4.170), (4.171) of \reference{Avramidi:1986mj}.} 
to the action, \eqn{einsteinframe2}. We must have $a,b>0$ and 
\beq
\mathrm{tensor}{:}\ \frac{2b}{3a}<1{+}\frac{3\xi^2a}{h_3};\quad 
\mathrm{scalar}{:}\ 18\xi^2<\frac{h_3}{b}.
\eeq
This calculation has been regarded as purely classical. When quantum 
corrections are calculated, these couplings become running couplings, 
and these inequalities must be respected at a certain 
symmetry-breaking scale $v$ that will be defined precisely in the next 
section. We wrote these inequalities in a form that takes advantage of 
the fact that $\xi(\mu)$ and the ratios of couplings $b(\mu)/a(\mu), 
h_3(\mu)/a(\mu)$ approach finite UVFPs as $\mu{\to}\infty,$ so it is most 
convenient to study their running in the range of scales above $v,$ as we did 
in \reference{Einhorn:2016mws}. We now turn to the determination of the 
quantum corrections to the effective action.

\section{Scale of Symmetry Breaking and the Dilaton Mass}\label{sec:dilaton}

The developments in the preceding sections all stemmed from the supposition that 
\beq\label{eq:vevPhi}
\vev{\Phi}=\vev{\Omega}\vev{\vsig}\neq0,
\eeq
which permitted transformation to the Einstein frame. In that frame, unlike the 
Jordan frame, we were able to identify the classical values of the Planck 
mass, $M_P,$ the cosmological constant $\Lambda,$ $\vev{\vsig}, \vev{\wt{R}}$ 
as well as the masses for all the fields except for the dilaton $\zeta,$ which 
classically appears as a free, massless scalar\footnote{In the Jordan frame, 
had we made the assumption \eqn{eq:vevPhi}, we \emph{could} have 
performed an expansion about that background, but it is still simpler to 
do in the Einstein frame with no nonminimal coupling(s) to $R.$}. In this 
section, we wish to determine $\vev{\zeta},$ or equivalently $\vev{\Omega},$ 
and the dilaton mass $m_d,$ by giving the radiative corrections to the 
effective action.

In general, the analytic calculation of radiative corrections to the effective 
action is impossible, and it has seldom been done for any 
spacetime-dependent background $\vev{R(x)}$ or $\vev{\Phi(x)}.$ (This is 
also true for fields in flat spacetime, with instantons being an 
exception~\cite{tHooft:1976snw}.) For backgrounds having $\vev{R}$ and 
$\vev{\Phi}$ spacetime independent, the one-loop corrections can be 
performed; even then, only bits and pieces of the two-loop corrections have 
been calculated to date.

Turning to the dilaton field $\zeta$, the classical action \eqn{einsteinframe} 
depends on $\zeta$ only through its gradient $\nabla_\mu\zeta=\pa_\mu\zeta,$
reflecting its role as a Goldstone boson associated with scale breaking 
$\vev{\Omega}\neq0.$ As remarked earlier, classical scale invariance is 
explicitly broken in the QFT, and the dilaton will get a nonzero mass $m_d$ 
at two-loop order. To determine whether or not it represents an instability, we 
shall have to calculate these radiative corrections. As mentioned in 
\secn{sec:jordanreal}, in our earlier work~\cite{Einhorn:2016mws} in the 
Jordan frame, we showed that the two-loop corrections responsible for 
$m_d\neq0$ could be calculated knowing only the one-loop $\beta$-functions. 
We also learned that $m_d^2>0$ for some range of couplings, but, being 
unsure of the proper normalization of the dilaton field, we could only determine 
$m_d$ within a multiplicative factor. Here, we wish to confirm those results and 
to determine the dilaton mass $m_d$ more precisely.

In order to be able to compare with our previous 
work~\cite{Einhorn:2016mws}, we begin with the form given there for the 
effective action in the Jordan frame in dS background: 
\beq\label{eq:effactJ}
\frac{\Gamma^{(J)}}{V_4}{=}\frac{S_{cl}^{(J)}(r)}{V_4}{+}
\frac{1}{2}B(r)\log\frac{R_J}{\mu^2}{+}
\frac{1}{8}C(r)\left(\!\log\frac{R_J}{\mu^2}\!\right)^{\!2}+\ldots,
\eeq
where, we recall, the ratio $r\equiv T_2/R_J,$ and $V_4$ has been defined 
earlier below \eqn{volume}. This presumes that both $\Phi$ and $R_J$ are 
constant. For constant $\Phi$, the transformation from Jordan 
to the Einstein frame yields $R_J {\to} \Omega^2 \wt{R}$ and 
$r {\to}M^2/\wt{R},$ independent of $\Omega.$ With $\rho\equiv(\wt{R})^{1/2},$ 
defined beneath \eqn{volume},
\beq\label{eq:RJtoRE}
\log\frac{R_J}{\mu^2} \to 
2\left(\frac{\zeta}{\sqrt{6\xip}M}{+}\log\frac{\rho}{\mu}\right).
\eeq
In the Jordan frame, we thought of the one-loop corrections as bringing in the 
dependence on the scalar curvature $R_J$ for a fixed ratio $r.$ By contrast,
in the Einstein frame, fixed $r$ represents fixed scalar curvature, $\wt{R},$ 
and the dependence on the dilaton field $\zeta$ enters through the 
corrections.

The effective action, like the classical action, is dimensionless, so it is not 
rescaled or changed by a conformal transformation. Therefore, 
\eqn{eq:effactJ} becomes 
\begin{align}\label{eq:effact3}
\begin{split}
\frac{\Gamma^{(E)}}{V_4} &{=}
\frac{S_{cl}^{(E)}\!\Big(\!\displaystyle{\frac{M^2}{\rho^2}}\Big)\!}{V_4}{+}
B\bigg(\!\frac{M^2}{\rho^2}\bigg)\!\bigg(\!\frac{\zeta}{\sqrt{6\xip}M}{+}
\log\frac{\rho}{\mu}\!\bigg){+}\!\\
&\hskip10mm \frac{1}{2}C\bigg(\!\frac{M^2}{\rho^2}\bigg)\!
\bigg(\!\frac{\zeta}{\sqrt{6\xip}M}
{+}\log\frac{\rho}{\mu}\bigg)^{\!\!2}\!+\ldots.
\end{split}
\end{align}

The original scale invariance is still reflected indirectly in \eqn{eq:effact3} by 
the property that, in the parentheses involving $\log(\rho/\mu)$, changing
the normalization scale $\mu$ can be offset by a shift in $\zeta.$ 
Thus, although $\zeta$ is no longer derivatively coupled when radiative
corrections are included, its value $\vev{\zeta}$ is not renormalization group
invariant, and, therefore, not directly observable. We shall exploit this shift 
freedom shortly.

All dependence on $\zeta$ in the effective action, \eqn{eq:effact3}, 
enters through the radiative corrections. The first derivative is 
\beq\label{eq:partialzeta1}
\begin{split}
\frac{1}{V_4}\frac{\pa \Gamma^{(E)}}{\pa\zeta} &=
\left(\!\frac{1}{\sqrt{6\xip}M}\right)\left[B\!\left(\!\frac{M^2}{\rho^2}\right)+
\right.\\
&\left.\hskip4mm C\!\left(\!\frac{M^2}{\rho^2}\right)
\left(\!\frac{\zeta}{\sqrt{6\xip}M}{+} \log\frac{\rho}{\mu}\!\right)\right].
\end{split}
\eeq

To one-loop order, $B{\to}B_1$ and $C{\to}0.$ To have an extremum in $\zeta,$ 
therefore, it must be that $B_1(M^2/\rho^2)=0.$ To this order, we may replace 
$\rho$ by its classical value, $\rho_0{=}\sqrt{h_3/(6\xi)}M$ from \eqn{eq:rho0},
so the extremum in $\zeta$ is determined by the 
equation
\beq\label{dtscale}
B_1\big(6\xi(\mu)/h_3(\mu)\big)=0.
\eeq
This equation is not true for all choices of $\mu.$ Its interpretation is that, in 
order for perturbative DT to occur, we must be able to find a scale, $\mu=v,$ 
at which this relation among couplings holds. In previous work in Jordan 
frame~\cite{Einhorn:2016mws}, we have obtained an explicit formula for 
$B_1(r),$ and \eqn{dtscale} is in fact identical to the Jordan frame condition for an 
extremum in $R_J$ at fixed $r.$ We showed that this equation can be satisfied for a 
range of coupling constants within the basin of attraction of the UVFP. 

Because \eqn{dtscale} is independent of $\zeta,$ we must go beyond 
one-loop order to determine a nonzero mass for the dilaton. Even at one-loop 
order, however, we expect the classical minimum in the curvature at 
$\rho_0$ to change slightly, $\rho_0{\to}\rho_0{+}\delta\rho_0,$ to which end we 
calculate the first derivative of the effective action with respect to $\rho\!:$\!
\begin{align}\label{eq:gammarho}
\begin{split}
\frac{1}{V_4}&\frac{\pa \Gamma^{(E)}}{\pa \rho}=
\frac{1}{V_4}\frac{\pa S_{cl}^{(E)}}{\pa \rho}+
\frac{1}{\rho}B\!\left(\!\frac{M^2}{\rho^2}\right)\\
&-\frac{2M^2}{\rho^3}B'\!\left(\!\frac{M^2}{\rho^2}\right)\!
\left(\!\frac{\zeta}{\sqrt{6\xip}M}\!+
\log\frac{\rho}{\mu}\!\right){+}\! \\
&\frac{1}{\rho}C\!\left(\!\frac{M^2}{\rho^2}\right)\!
\left(\!\frac{\zeta}{\sqrt{6\xip}M}\!+
\log\frac{\rho}{\mu}\!\right)+\ldots,
\end{split}
\end{align}
where we truncated the equation for reasons to be explained below. As with 
\eqn{eq:partialzeta1}, the one-loop correction has $B{\to}B_1$ and $C{\to}0.$ 
It is convenient to choose the normalization scale $\mu=v,$ at which 
\eqn{dtscale} holds, so that the second term on the RHS in 
\eqn{eq:gammarho} vanishes. Then, to first order, we expand in $\delta\rho_0$ 
to get
\begin{subequations}\label{eq:oneloopdeltarho}
\begin{align}\label{eq:gammarho1}
\begin{split}
&\frac{1}{V_4}\frac{\pa \Gamma^{(E)}}{\pa \rho}\approx 
\frac{1}{V_4}\frac{\pa^2 S_{cl}^{(E)}}{\pa\rho^2}\Bigg|_{\rho_0}\!\delta\rho_0-\\ 
&\hskip8mm\frac{2M^2}{\rho_0^3}B_1'\!\left(\!\frac{M^2}{\rho_0^2}\right)
\bigg(\!\frac{\zeta}{\sqrt{6\xip}M}\!+\log\frac{\rho_0}{v}\!\bigg),
\end{split}\\
\label{eq:deltarho0}
&\hskip5mm {\approx} \frac{2M^2}{\rho_0^3}\!\bigg[
\frac{\xi\delta\rho_0}{\rho_0}{-} B_1'\!\bigg(\!\frac{M^2}{\rho_0^2}\bigg)\!\bigg(\!\frac{\zeta}{\sqrt{6\xip}M}
{+}\log\frac{\rho_0}{v}\!\bigg)\!\bigg]\!.\!
\end{align}
\end{subequations}
Given $M,$ setting \eqn{eq:deltarho0} to zero and $\zeta{\to}\vev{\zeta}$ 
determines a relation between $\delta\rho_0$ and $\vev{\zeta}.$ The value 
of $\vev{\zeta}$ is still not fixed at one-loop order, as we pointed out earlier. 
Therefore, we may conveniently choose $\vev{\zeta}$ such that 
\beq\label{vevzeta}
\vev{\zeta}{+}\sqrt{6\xip(v)}M\log\frac{\rho_0}{v}= 0,
\eeq
where $\xip(v)\equiv\xi(v){+}1/6,$ and $\rho_0$ is given in \eqn{eq:rho0}. 
With this choice for $\vev{\zeta},$ we may conclude that the first-order 
correction $\delta\rho_0$ vanishes!

Although we have a one-loop constraint \eqn{dtscale} consistent with
$\zeta$ having a local extremum, we have not determined its character.
To do so requires going to two-loop order. Although not all two-loop 
corrections to the effective action or to the $\beta$-functions are known, some 
two-loop effects are calculable from the one-loop 
$\beta$-functions~\cite{'tHooft:1973mm}, including $C_2(r),$ the first nonzero 
contribution to $C(r).$ Fortunately, these turn out to be sufficient to determine 
the two-loop contributions to the effective action that are 
required~\cite{Einhorn:2016mws}. 

To see that in the present language, we need the second variations which, 
on-shell with our conventions for $M$ and $\vev{\zeta}$, take the form
\begin{subequations}
\begin{align}
\frac{1}{V_4}\frac{\pa^2 \Gamma^{(E)}}{\pa\zeta^2}\bigg|_{os}
&\hskip-0.1in=
\frac{1}{{6\xip}M^2}C_2\!\left(\!\frac{M^2}{\rho_0^2}\right),
\label{eq:partialzeta2}\\
\begin{split}
\frac{1}{V_4}\frac{\pa^2 \Gamma^{(E)}}{\pa\zeta\pa\rho}\bigg|_{os}&\hskip-0.1in=
\left(\!\frac{1}{\sqrt{6\xip}M}\right)\! 
\bigg[{-}\frac{2M^2}{\rho_0^3}B_1'\!\left(\!\frac{M^2}{\rho_0^2}\right){+}\!\\
&\hskip5mm \frac{1}{\rho_0}C_2\!\left(\!\frac{M^2}{\rho_0^2}\right)\bigg],
\label{eq:partialzetarho}
\end{split}\\
\begin{split}
\frac{1}{V_4}\frac{\pa^2 \Gamma^{(E)}}{\pa\rho^2}\bigg|_{os}&\hskip-0.1in=
\frac{1}{V_4}\frac{\pa^2 S_{cl}^{(E)}}{\pa{\rho^2}}\bigg|_{\rho=
\rho_0}\hskip-0.12in
-\frac{2}{\rho_0^3}B_1'\!
\left(\!\frac{M^2}{\rho_0^2}\right)+\\
&\hskip5mm \frac{1}{\rho_0}C_2\!\left(\!\frac{M^2}{\rho_0^2}\right)+\ldots,
\label{eq:partialrho2}
\end{split}
\end{align}
\end{subequations}
where the subscript ${os}$ refers to the value ``on-shell." 
Having arranged for the one-loop correction to $\rho_0$ to vanish, 
this means $\rho{\to}\rho_0, \zeta{\to}\vev{\zeta},$ with $\vev{\zeta}$ given 
by \eqn{vevzeta}. In the last line, \eqn{eq:partialrho2}, we have omitted certain 
other one- and two-loop contributions for reasons that will become clear 
shortly. 
Consider the matrix of second variations
\beq
\label{eq:delta2onshell}
\delta^{(2)}\Gamma^{(E)}{=}\frac{1}{2}
\begin{pmatrix}
 \delta\zeta \!&\! \delta\rho
\end{pmatrix}
\!\!
\begin{bmatrix}
\displaystyle{ \frac{\pa^2 \Gamma^{(E)}}{\pa\zeta^2}} &
 \displaystyle{\frac{\pa^2 \Gamma^{(E)}}{\pa\zeta\pa\rho}}\smallskip\\
\displaystyle{\frac{\pa^2 \Gamma^{(E)}}{\pa\zeta\pa\rho}}&
\displaystyle{\frac{\pa^2 \Gamma^{(E)}}{\pa\rho^2}}
\end{bmatrix}
\!\!
\begin{pmatrix}
\delta\zeta \\ \delta\rho
\end{pmatrix}\!.\!
\eeq
To review the order of the matrix elements, we recall that the leading 
nonvanishing term of \eqn{eq:partialzeta2} is $O(\hbar^2),$ \ie two loops; of
\eqn{eq:partialzetarho}, $O(\hbar);$ of \eqn{eq:partialrho2}, $O(1).$ Thus, the 
matrix has a familiar ``see-saw" pattern, the same structure that was 
encountered in the Jordan frame calculation~\cite{Einhorn:2016mws}. The 
determinant is $O(\hbar^2)$ and the trace is $O(1),$ so one eigenvalue is 
$O(1)$ and the other $O(\hbar^2).$ 

Naturally, the larger one is associated with the classical 
fluctuation determined in the previous section. To be precise, we take 
the classical approximation for ${\pa^2 \Gamma^{(E)}}/{\pa\rho^2}$, 
\viz $2\xi M^2/\rho_0^4$ from just below \eqn{eq:rho0}. Then 
the larger eigenvalue of the matrix in \eqn{eq:delta2onshell}
\beq\label{largeeps}
\varepsilon_1=\frac{2\xi M^2 V_4}{\rho_0^4}+O(\hbar^2),
\eeq
with eigenvector $(\delta\zeta,\delta\rho){=}(0,1){+}O(\hbar^2).$
When divided by the spacetime volume $V_4/\rho_0^4,$ $\varepsilon_1$ 
gives precisely the value $m_\rho^2$ in \eqn{rhomass}. Having arranged for 
the one-loop correction to $m_\rho$ to vanish, we are not really interested in 
its two-loop corrections. In fact, they would require 
$B_2,$ which is not known and cannot be determined using one-loop 
$\beta$-functions. It would also require taking into account gravitational 
corrections to the wave-function renormalization.

The smaller eigenvalue $\varepsilon_2$ is associated with the dilaton, 
\beq\label{smalleps}
\varepsilon_2=\Big(\frac{1}{16\pi^2}\Big)^{\!2} \frac{V_4}{6\xip M^2}
\left[C_2-\frac{B'^2_1}{2\xi}\right]+O(\hbar^3),
\eeq
with eigenvector 
\beq\label{mixing}
(\delta\zeta,\delta\rho){=}(1,{-}\sqrt{h_3/(\xip\xi^3)\!}B'_1/6){+}O(\hbar^2).
\eeq
All scale-dependent quantities on the RHS of \eqns{smalleps}{mixing}
are to be evaluated at the DT scale $\mu\!=\!v,$ where \eqn{dtscale} is fulfilled.
We have made explicit the factors of $16\pi^2,$ heretofore suppressed, in 
order to emphasize how very much smaller than $\varepsilon_1$ 
this is. If we divide by the spacetime volume, we find 
\beq\label{dilatonmass2}
\frac{m_d^2}{M_P^2}=\!\bigg(\!\frac{1}{16\pi^2}\!\bigg)^{\!\!2}
\frac{1}{30\xi\xip}\bigg(\!\frac{h_3}{6\xi}\!\bigg)^{\!\!2}
\Bigg[C_2{-}\frac{B'^2_1}{2\xi}\Bigg]\Bigg|_{\mu=v},
\eeq
corresponding to a term in the effective action 
\beq\label{eq:mdsquare}
\int\! d^4x \sqrt{\wt{g}}\; \frac{m_d^2}{2}\, \delta\zeta^2
{+}O(\hbar^3).
\eeq
Are we really justified in identifying this with the dilaton mass? We believe the 
answer is yes, although it does require further justification. The fact that 
the eigenvalue \eqn{smalleps} is already of $O(\hbar^2)$ hides a multitude of 
sins of omission. For example, we did not address the one-loop corrections to 
the spacetime volume\footnote{Of course, we could have raised the same 
point below \eqn{largeeps}. Fortuitously, having arranged in \eqn{vevzeta} for 
$\delta\rho_0\!=\!0$ in $O(\hbar),$ such corrections to \eqn{largeeps} will be at 
least $O(\hbar^2).$}, but that is clearly not necessary in order to determine 
$\varepsilon_2$ through $O(\hbar^2)$ in \eqn{smalleps}. Similarly, any mixing of 
$\delta\zeta$ with $\delta\rho$ affects $m_d^2$ in $O(\hbar^3)$ or higher.

There is also the related issue of whether the kinetic term (wave function 
normalization) for $\delta\zeta$ is canonical. The preceding calculations 
assumed the fluctuations $(\delta\zeta,\delta\rho)$ were constant, but we must 
go beyond the static limit to answer this question. In fact, a glance at the 
original Einstein-frame action \eqn{einsteinframe} casts doubt on this. Besides 
the canonical term for $\zeta$ in \eqn{eq:emat}, there are also the 
higher-derivative terms in \eqn{eq:esho}. Since they are classical, \ie $O(\hbar^0),$ 
they cannot be ignored in general. 

We propose to deal with them as follows:  As will be discussed further in 
\secn{sec:left}, below the Planck scale $M_P,$ the gravitational theory is 
well-approximated by the E-H action plus higher-dimensional operators.  
This is in effect a derivative expansion in $1/M_P.$  Since $m_d\ll M_V\lesssim v,$ 
we may consider an expansion in $1/v$  on momentum scales small compared to all 
the particles that acquire masses after SSB in tree approximation in Einstein frame, 
which were discussed in 
Sections~\ref{sec:vbmass}, \ref{sec:scalarmass}, and \ref{sec:rho}. In that case, all 
the terms in \eqn{eq:esho} involving $\vartheta_\mu$ comprise operators of higher 
dimension than four and, thus, will be small compared with the terms remaining. 
Then the leading contribution to the kinetic term for 
$\zeta$ is entirely from \eqn{eq:emat}, which is simply the canonical term 
$(\wt{\nabla}\zeta)^2/2.$ In that case, what we have called $m_d^2$ in 
\eqn{dilatonmass2} above is in fact correct. Furthermore, since $-\wt{\Box}$ 
is an elliptic operator on Hilbert space, we may conclude that the non-static 
fluctuations in this approximation will also be stable.

\section{Low-energy Effective Field Theory}\label{sec:left}

There have been several physical scales variously identified as 
$M_P, \Lambda, M_V, M_{\varDelta\vsig}, m_\rho,$ as well as 
$v$ and $m_d,$ their ratios are in principle observables. Unfortunately, 
all but $m_d,$ are likely to be $O(M_P),$ although we have not 
exhaustively explored the range of parameter space delineated by
the determination of $v,$ \eqn{dtscale}, and the requirement that 
$\varepsilon_2>0,$ \eqn{smalleps}, or, equivalently, $m_d^2>0,$ 
\eqn{dilatonmass2}. As with superstring phenomenology, the only natural realm of 
application at such scales is to precision cosmology around the time of the Big Bang 
and earlier. On the other hand, unlike superstring theory, QFT can deal with the time 
evolution of (gauge-invariant) correlation functions, provided the measurement 
frame is specified. This calls attention to the issue of whether renormalizable gravity 
is unitary at scales above $v.$ We expect to have more to say about this in the 
future, but we will have little to contribute to the debate in this paper. 

Near the end of the previous section, we argued that there may be a range of 
momentum scales, $m_d<p\lesssim M_V\lesssim M_P$ in which all particles 
except the massless vector bosons of $\su51,$ the massless graviton, and 
the dilaton have become irrelevant. The corresponding low-energy, classical 
action can be extracted from \eqn{einsteinframe} with the inclusion of 
the dilaton mass term \eqn{eq:mdsquare}:
\begin{align}\label{eq:left1}
\begin{split}
S^{(E)}_{eff}&=\!\!\int\!\! d^4x\sqrt{\wt{g}} \Bigg[\frac{1}{2a}\wt{C}^2{+}
\frac{1}{3b}\wt{R}^{2}+c\,\wt{G}+\frac{h_3}{24}M^4\\
&\hskip2mm 
{-}\frac{\xi M^2 }{2} \wt{R}+
\frac{1}{4}\Tr[\wt{F}_\mn^2]{+}\frac{(\wt{\nabla}\delta\zeta)^2}{2}
+\frac{m_d^2\,\delta\zeta^2}{2}\Bigg],\!
\end{split}
\end{align}
where $\wt{F}_\mn$ represents the $\su51$ field strength for the massless 
gauge bosons. Recalling \eqns{eq:zeta}{mplanck}, we have neglected terms 
involving $\theta_\mu=\sqrt{\xi/6\xip}\,\wt{\pa}_\mu\zeta/M_P,$ with 
$\xip=\xi{+}1/6,$ since, for this range of energy scales, these terms are 
of the same order as others dropped. Only the dilaton and the 
massless vectors of $\su51$ remain in addition to the metric 
$\wt{g}_\mn.$ We could also calculate some of the higher-dimensional 
operators that have been neglected, but they are not of great interest for 
present purposes unless the low-energy, effective action based on 
\eqn{eq:left1} proved to be unstable or to have zero-modes that may be 
removed by such higher-order terms.

The on-shell solution for the background turns out to correspond to constant  
curvature $\wh{R}=4\Lambda,$ with $\Lambda=h_3 M_P^2/(24\xi^2).$ 
Assuming that the background has dS global symmetry, as in our earlier 
discussions, we can expand about the background to explore stability. It will come 
as no surprise that the fluctuations will be stable, since we require $m_d^2>0.$ 
There will remain the by-now familiar five conformal, zero modes associated with 
coherent fluctuations about $S^4$ background. 

\section{Conclusions}\label{sec:conclude}

In \reference{Einhorn:2016mws}, we discussed a classically
scale-invariant model in which renormalizable gravity is coupled to
matter in the form of an $SO(10)$ gauge field plus a real scalar field
in the adjoint representation. We showed that the model contains a
locally stable UVFP, so that all couplings are AF. Moreover, the domain
of attraction of the UVFP includes a region of parameter space
corresponding to spontaneous breaking of the gauge symmetry to
$SU(5)\otimes U(1)$, with the scalar multiplet acquiring a VEV. This
VEV is perturbatively determined and calculable by DT, which determines
the scale at which a specific relationship among the various
dimensionless couplings holds true, and its presence generates an
E-H term from the nonminimal scalar coupling to gravity. The
quartic behavior of the metric's propagator may not admit an ordinary
particle interpretation, but it is not an obstacle to the calculation 
of Euclidean correlation functions. 

In this paper, the same model was transformed from Jordan to the Einstein frame, 
and the form of the one-loop effective action there was further developed. The 
Planck mass, cosmological constant, vector boson masses, and related scales 
were unambiguously identified. In particular, we were able to identify the 
canonically normalized dilaton field $\zeta$ and to determine the dilaton-mass, 
$m_d,$ \eqn{dilatonmass2}. We wish to re-emphasize that, even though 
the mass is a two-loop effect, it can be calculated knowing only the one-loop 
beta-functions. Whether such a light dilaton plays an important role in 
cosmological applications remains to be determined in future, as do other 
issues such as inflation and Dark Energy.

We showed that the effective field theory below the scale of symmetry
breaking takes the form of the gauged $\su51$ nonlinear
sigma model plus a dilaton and graviton. Of course, we would like this to 
be prototypical of a realistic model; obviously much remains to be done 
with regard to demonstrating a realistic SM-like theory at low energies, 
including in particular the emergence of the electroweak scale.

Although we have likened our determination of the symmetry-breaking
scale dynamics to DT \`a~la Coleman-Weinberg~\cite{Coleman:1973jx}, we
wish to reemphasize certain differences from their mechanism. In their
seminal treatment, the self-coupling $\lambda(\mu)$ of the scalar field
is unusually small in a neighborhood of the DT scale $\mu=v.$ Indeed,
$\lambda(v)$ is of the same order as the one-loop amplitude,
$O(\alpha^2),$ very near to where $\lambda(\mu)=0.$ 

In our application, the picture is different and is in fact frame
dependent.  In the Jordan  frame~\cite{Einhorn:2016mws}, which is most
nearly similar to  \reference{Coleman:1973jx}, we first determined the
direction of symmetry-breaking and  the ratio $\vev{\Phi}/\rho,$ where
$\rho\equiv\sqrt{\vev{R_J}},$ from extremizing the  classical potential.
We then determined the value of the scalar curvature $\rho=v$ from  the
radiative corrections.  In a neighborhood of this scale $v,$ the
one-loop correction to the effective  action $\Gamma^{(J)},$
\eqn{eq:effactJ}, becomes unusually small, of order of  the two-loop
correction. More precisely, we seek the value of $\rho$ at which 
\beq\label{eq:j1}
\rho\frac{\pa\Gamma^{(J)}}{\pa\rho}\bigg|_{\rho=v}\!\!
=B_1(\mu){+}B_2(\mu)+C_2(\mu)\log\frac{v}{\mu}=0.
\eeq
If we choose the normalization scale $\mu=v,$ \eqn{eq:j1} simplifies to 
$B_1(v){+}B_2(v)=0.$ Thus, at the extremum, the one-loop correction $B_1$ is 
of order of the two-loop correction $B_2(v).$ In first approximation, the 
extremum occurs where $B_1(v)=0,$ a relation among couplings at scale $v$.
In short, as compared with DT in \reference{Coleman:1973jx}, our application is 
of higher order in the loop-expansion. Instead of the extremum occurring at the 
scale $v$ where a tree coupling $\lambda$ falls to $O(\hbar)$ corrections, our 
extremum $v$ is determined by the scale at which the $O(\hbar)$-correction 
falls to $O(\hbar^2).$ The determination that the extremum is in fact a minimum 
is a two-loop effect which, fortunately, was calculable from the one-loop 
$\beta$-functions.
 
In the Einstein frame, \eqn{einsteinframe}, the story was rather different, although the 
results were the same. Since only the $\Omega(x)$ field was 
used in performing the conformal transformation, the calculation in 
\secn{sec:left} of the ``directions" $\vev{\vsig}$ at which extrema occur 
was essentially the same as before in \reference{Einhorn:2016mws}, as was the 
determination of which one was a local minimum. Unlike the Jordan frame
calculation, we were able to determine a first approximation to the scalar 
curvature $\rho\equiv(\wt{R})^{1/2}$ and to show it was a local minimum 
already in tree approximation, \eqn{eq:rho0}. In contrast, the dilaton degree of 
freedom $\zeta$ enters the effective potential \eqn{eq:effact3} only via 
radiative corrections, and, we found the DT equation $B_1(v)=0$ as a result of 
seeking the extremum in $\zeta.$ We were also able to calculate the one-loop 
correction $\delta\rho_0$ to the curvature $\rho_0,$ \eqn{eq:deltarho0}, and, 
by a propitious choice for $\vev{\zeta},$ we arranged for it to vanish. We then 
were able to calculate the curvature in $\zeta$ and thereby determine the 
dilaton mass $m_d,$ \eqn{dilatonmass2}, something which we had only be 
able to estimate previously. 

In the original description in the Jordan frame, it was clear that the metric has 
a scalar DoF, \ie that this is a scalar-tensor theory of gravity. In the Einstein 
frame, this DoF was represented by the conformal field $\rho$ in 
\secn{sec:dilaton}. In the low energy effective field theory, \secn{sec:left}, this 
scalar DoF does not 
appear, \ie it decouples (except for the five zero modes.) Even though the 
dilaton mass is proportional to the scale of SSB, it is a two-loop effect and 
$m_d^2/M_P^2\ll 1,$ \eqn{dilatonmass2}. Unlike the other massive 
scalars, it does involve mixing with the scalar DoF of the metric. 

These conclusions do not depend in detail on this particular model, and we 
expect them to be generic. While that is hopeful for finding a renormalizable 
extension of the SM to include gravity, it also suggests that it may be very 
difficult to test experimentally.

We have not discussed analytic continuation from Euclidean to Lorentzian
signature. We simply assumed that for relevant spacetimes, it can be 
performed. New issues arise however: not even dS remains compact, 
although, depending on the frame, it is often the case that a fixed time
slice has  compact spatial volume. Although correlation functions remain
well-defined, they can become IR divergent as the timelike separation
between spacetime  points grows indefinitely.  This further
complicates the discussion of unitarity, but in the past, all such
perturbative infrared divergences in  QFT have been resolved by a
careful specification of observables. Regardless,  having settled the
primary issues of instability and ghosts that caused this line of 
investigation to be abandoned nearly 40 years ago, we are optimistic
that  eventually  asymptotically free models based on renormalizable
gravity will  turn out to be consistent, unitary completions of
Einstein-Hilbert gravity. Whether  they can be extended to include the
SM fields while preserving naturalness down to the electroweak scale
remains a theoretical challenge. 

\subsection{{\bf Post-publication note added}}\label{note:added}%\hyperref[note:added]{{\bf Note added:}}}
\vskip-3mm 
\noindent In connection with our remarks about $f(R)$ models related to  footnote \ref{footnote:reviews}, we inadvertently forgot to include \reference{Alvarez-Gaume:2015rwa}, which analyzes models quadratic in curvature invariants, with particular emphasis on $R^2$ models both with and without an E-H term and a cosmological constant. We agree with these authors  concerning the relation of the signs of the couplings $a, b$ to ghosts and tachyons as well as the sign of the E-H term (set by $\xi$) for attractive gravity. Their analysis is essentially classical and, because $f(R)$ models are not renormalizable, not even as effective field theories, we do not entirely agree with their view of the UV/IR relationship. In particular, we restrict our attention to AF models and do not turn to string theory for a completion of quantum gravity.

\acknowledgments
This research was supported in part by the National Science Foundation under
Grant No. PHY11-25915 (KITP) and by the Baggs bequest (Liverpool). 
One of us (MBE) would like to thank Arkady Tseytlin for discussions.

\appendix

%%%%%%%%%%
\section{Conformal Redefinition of Metric}\label{sec:conftransf}
%%%%%%%%%%

In this Appendix, we summarize some formulas associated with conformal
transformations of the metric\footnote{With appropriate adjustments for 
sign conventions, our formulae agree with Appendix~G of 
\reference{Carroll:2004st}.}. As in the text, we choose Euclidean
signature and choose the definition of the Ricci tensor so that $R>0$
for positive curvature. Given the definitions, \eqn{eq:conftransf}, 
$\widetilde{g}_\mn\equiv\Omega^{-2} g_\mn,$ then obviously
$\sqrt{\widetilde{g}}=\Omega^{-4}\sqrt{{g}},$ or in $n$-dimensions,
$\sqrt{\widetilde{g}}=\Omega^{-n}\sqrt{{g}}.$ 

Relations among conformally-related curvatures are
often more simply expressed when written in terms of
$\varUpsilon\equiv\log(\Omega)$, and here we state the results in terms
of $\varUpsilon$ instead of $\Omega.$

Some useful identities are
\begin{subequations}
\begin{align}
\Omega^{-1}\nabla_\mu\Omega&=\nabla_\mu \varUpsilon,\\
\Omega^{-1}\nabla_\mu\nabla_\nu\Omega
&=\nabla_\mu \varUpsilon\;
\nabla_\nu \varUpsilon+\nabla_\mu\nabla_\nu \varUpsilon,\\
\Omega\nabla_\mu\nabla_\nu\Omega^{-1}\!
&=\nabla_\mu\varUpsilon\;\nabla_\nu\varUpsilon
-\nabla_\mu\nabla_\nu\varUpsilon.
\end{align}
\end{subequations}
The conformal transform of the connection is
\begin{subequations}
\begin{align}
\Gamma^\kappa_\mn&=g^\kl(\pa_\mu g_{\nu\lambda}+
\pa_\nu g_{\mu\lambda}-\pa_\lambda g_\mn)/2\to\\
\begin{split} 
 \widetilde{\Gamma}^\kappa_\mn&=\Gamma^\kappa_\mn{-}
\Delta^\kappa_\mn,\ {\rm with}\\ 
\Delta^\kappa_\mn&\equiv\delta^\kappa_\mu\pa_\nu\varUpsilon{+}
\delta^\kappa_\nu\pa_\mu\varUpsilon
{-}g_\mn g^{\kappa\lambda}\pa_\lambda\varUpsilon.\!
\end{split}
\end{align}
\end{subequations}
$\Delta^\kappa_\mn$ transforms as an ordinary tensor under general 
coordinate transformations. The Riemann curvature,
$R^\kappa{}_{\lambda\mn}\!=
\pa_\mu\Gamma^\kappa_{\lambda\nu}{-}\pa_\nu\Gamma^\kappa_{\lambda\mu}
{+}\Gamma^\kappa_{\mu\rho}\Gamma^\rho_{\lambda\nu}{-}
\Gamma^\kappa_{\nu\rho}\Gamma^\rho_{\lambda\mu},$ 
transforms as 
\begin{align}\label{eq:riemanntrans}
\begin{split}
\widetilde{R}^\kappa{}_{\lambda\mn}&=R^\kappa{}_{\lambda\mn}
+\delta^\kappa_{[\mu}\nabla_{\nu]}\vartheta_\lambda+
\nabla^\kappa\vartheta_{[\mu}g_{\nu]\lambda}
+\\
&\hskip10mm \delta^\kappa_{[\mu}\vartheta_{\nu]}\vartheta_\lambda
-\vartheta^2\delta^\kappa_{[\mu} g_{\nu]\lambda},
\end{split}
\end{align}
where $\vartheta_\mu\equiv\pa_\mu\varUpsilon,$ 
$\nabla_\mu\vartheta_\nu\equiv\pa_\mu\vartheta_\nu-
\Gamma^\lambda_\mn\vartheta_\lambda.$
Thus,
\begin{subequations}
\begin{align}\label{eq:riccitrans}
\begin{split}
\widetilde{R}_\mn&\!=R_\mn+(n-2)\nabla_\mu\vartheta_\nu+
\left(\nabla\!\!\cdot\!\vartheta\right)g_\mn\\
&+
(n-2)\left(\vartheta_\mu\vartheta_\nu-\vartheta^2g_\mn\right),
\end{split}\\
\label{eq:tfriccitrans} \widetilde{\widehat{R}}_\mn&\!=\widehat{R}_\mn{+}
(n{-}2)\!\Big[\nabla_\mu\vartheta_\nu{+}\vartheta_\mu\vartheta_\nu
{-}\frac{g_\mn}{n}\!\big(\nabla\!\!\cdot\!\vartheta 
{+} \vartheta^2\big)\!\Big],\!\!\\
\label{eq:scalartrans}
\widetilde{R}&=\Omega^2\left[R
+2(n-1)\nabla\!\!\cdot\!\vartheta -(n-1)(n-2)\vartheta^2\right]\!.\!
\end{align}
\end{subequations}
Note that $\nabla\!\!\cdot\!\vartheta=\Box\varUpsilon.$ For $n=4,$ 
these become
\begin{subequations}
\begin{align}
\widetilde{R}_\mn&\!=R_\mn{+}2\nabla_\mu\vartheta_\nu{+}
\big(\nabla\!\!\cdot\!\vartheta\big)g_\mn{+}
2\big(\vartheta_\mu\vartheta_\nu{-}\vartheta^2g_\mn\big),\!\\
\widetilde{\widehat{R}}_\mn&=\widehat{R}_\mn{+}
2\Big(\nabla_\mu\vartheta_\nu{+}\vartheta_\mu\vartheta_\nu
{}\frac{g_\mn}{4}\big(\nabla\!\!\cdot\!\vartheta {+} \vartheta^2\big)\!\Big),\!\\
\widetilde{R}&=\Omega^2\left(R
+6\left(\nabla\!\!\cdot\!\vartheta -\vartheta^2\right)\right).
\end{align}
\end{subequations}

The Weyl tensor $C^\kappa{}_{\lambda\mn}$ is invariant under conformal 
transformations, so that $\sqrt{g}C^2$ is also invariant. 
Assuming that the conformal transform does not change the topology (i.e., Euler 
characteristic), then $G$ must change by $\nabla_\mu J^\mu$ for some current 
$J^\mu.$ Since $G=C^2-2W$, with $W\equiv\hat{R}_\mn^2-R^2/12,$ 
we find that $W$ transforms as
\bea\label{eq:wtrans}
\!\widetilde{W}\!\!&\!-\!&\!W = 4\nabla_\mu J^\mu,\ \ \mathrm{where}\\ 
\!J_\mu \!\!&\!\equiv\!&\! \vartheta_\nu\nabla_\mu\vartheta^\nu{-}
\vartheta_\mu \nabla\!\!\cdot\!\vartheta{+}\!
\left(\!R_\mn{-}g_\mn\frac{R}{2}\right)\vartheta^\nu{+}\vartheta^2\vartheta_\mu.\ \ 
\eea
To derive this result, we must calculate 
\begin{subequations}
\begin{align}
\sqrt{\widetilde{g}}\,\widetilde{W}
&=\sqrt{\widetilde{g}}\bigg(\widetilde{\widehat{R}}_\mn^2
-\frac{\widetilde{R}^2}{12}\bigg),\hskip30mm \\
\begin{split}
&=\sqrt{g}\bigg[\Big[\widehat{R}_\mn\!+
2\Big(\nabla_\mu\vartheta_\nu\!+\vartheta_\mu\vartheta_\nu-\\
&\hskip-2mm \frac{g_\mn}{4}\!\big(\nabla\!\!\cdot\!\vartheta
+\! \vartheta^2\big)\!\Big)\!\Big]^2\!-
 \frac{\left[R+6\left(\nabla\!\!\cdot\!\vartheta
-\vartheta^2\right)\right]^2}{12}\bigg].\!
\end{split}
\end{align}
\end{subequations}
Hence, letting $\Delta W\equiv \widetilde{W}-W,$ we find 
\begin{subequations}
\begin{align}
\begin{split}
\Delta W&=
4\widehat{R}^\mn\big(\nabla_\mu\vartheta_\nu+\vartheta_\mu\vartheta_\nu\big)
-R\left(\nabla\!\!\cdot\!\vartheta -\vartheta^2\right)+\\
&\hskip5mm 4\left(\nabla_\mu\vartheta_\nu+\vartheta_\mu\vartheta_\nu
-\frac{g_\mn}{4}\left(\nabla\!\!\cdot\!\vartheta 
+ \vartheta^2\right)\right)^2-\\
& \hskip5mm 3\left(\nabla\!\!\cdot\!\vartheta -\vartheta^2\right)^2,
\end{split}\\
\begin{split}
&=4\widehat{R}^\mn\nabla_\mu\vartheta_\nu{-}R\nabla\!\!\cdot\!\vartheta
{+}4\widehat{R}^\mn\vartheta_\mu\vartheta_\nu+R\vartheta^2{+}\\
&\hskip5mm 4\left(\nabla_\mu\vartheta_\nu+\vartheta_\mu\vartheta_\nu\right)^2
-\left(\nabla\!\!\cdot\!\vartheta + \vartheta^2\right)^2-\\
&\hskip5mm 3\left(\nabla\!\!\cdot\!\vartheta -\vartheta^2\right)^2,
\end{split}\\
\begin{split}
&=\big(4R^\mn\nabla_\mu\vartheta_\nu-2R\,\nabla\!\!\cdot\!\vartheta\big)+\\
&\hskip5mm 4\big(R^\mn\vartheta_\mu\vartheta_\nu+
\left(\nabla_\mu\vartheta_\nu\right)^2
-\left(\nabla\!\!\cdot\!\vartheta\right)^2\big)+ \\
&\hskip5mm 4\big(2\vartheta^\mu\vartheta^\nu\nabla_\mu\vartheta_\nu+
\vartheta^2\nabla\!\!\cdot\!\vartheta\big).
\label{eq:wchange}
\end{split}
\end{align}
\end{subequations}
In the last step, the squares were expanded into monomials and the 
various terms gathered into a polynomial in $\vartheta_\mu.$ For later
convenience, the terms involving the curvature were expressed in terms
of the usual Ricci tensor.

We now wish to show that the change $\Delta W$ can be written as 
the divergence of a vector. (Fortunately, the quartic terms 
involving $(\vartheta^2)^2$ canceled out in \eqn{eq:wchange}, as required.)
The linear terms may be written as 
$4\nabla_\mu\left[(R^\mn-g^\mn R/2)\vartheta_\nu\right],$ 
since the Einstein tensor has zero divergence. The cubic terms are also easily seen to be $4\nabla_\mu\left[\vartheta^2\theta^\mu\right].$ 

The quadratic terms require a bit more work. We will want to use the 
well-known relation\footnote{More generally, 
$[\nabla_\mu,\nabla_\nu]V_\lambda=R^\kappa{}_{\lambda\mn} V_\kappa$ 
which mathematicians~\cite{AlvarezGaume:1984dr} would write as 
$\nabla^2V=RV,$ the wedge product being understood. In this notation,
$\nabla^2\equiv\nabla\wedge\nabla\ne \Box.$} 
$R^\mn\vartheta_\nu=[\nabla^\nu,\nabla^\mu]\vartheta_\nu=
\nabla^\nu\nabla^\mu\vartheta_\nu-\nabla^\mu\nabla\!\!\cdot\!\vartheta,$ 
in order to write it as gradients like the other terms. 
We may take advantage of the fact that 
$\vartheta_\nu$ is itself the gradient of a scalar to rewrite 
$\nabla^\mu\vartheta_\nu
=\nabla^\mu\nabla_\nu\varUpsilon= \nabla_\nu\nabla^\mu\varUpsilon=
\nabla_\nu\vartheta^\mu,$ since two covariant derivatives commute when acting 
on a scalar. Hence, $R^\mn\vartheta_\nu=\Box\vartheta^\mu-
\nabla^\mu\nabla\!\!\cdot\!\vartheta.$ 

To bring all the quadratic terms into the form of a divergence, note that there are only 
two vector monomials that can be formed that are both quadratic in 
$\vartheta_\mu$ and have a single gradient, \viz 
$\vartheta_\nu\nabla_\mu\vartheta^\nu$ and $\vartheta_\mu\nabla\!\!\cdot\!\vartheta$, 
so the quadratic terms in the current $J^\mu$ must be a linear combination of these two 
vectors. Their divergences are
\begin{subequations}
\begin{align}
\nabla_\mu\left(\vartheta_\nu\nabla^\mu\vartheta^\nu\right)&=
\left(\nabla_\mu\vartheta_\nu\right)^2+\vartheta_\nu\Box\vartheta^\nu,\\
\nabla_\mu\left(\vartheta^\mu\nabla\!\!\cdot\!\vartheta\right)&=
\left(\nabla\!\!\cdot\!\vartheta\right)^2+
\vartheta^\mu\nabla_\mu\!\left(\nabla\!\!\cdot\!\vartheta\right).
\end{align}
\end{subequations}
Pulling all these pieces together, we find the quadratic terms become 
\begin{subequations}
\begin{align}
&\left[R^\mn\vartheta_\mu\vartheta_\nu+
\left(\nabla_\mu\vartheta_\nu\right)^2
-\left(\nabla\!\!\cdot\!\vartheta\right)^2\right]\\
\begin{split}
&=\vartheta_\nu\Box\vartheta^\nu-
\vartheta^\mu\nabla_\mu\!\left(\nabla\!\!\cdot\!\vartheta\right)+
\nabla_\mu\left(\vartheta_\nu\nabla^\mu\vartheta^\nu\right)-\\
&\hskip5mm \vartheta_\nu\Box\vartheta^\nu
 -\nabla_\mu\left(\vartheta^\mu\nabla\!\!\cdot\!\vartheta\right)+
 \vartheta^\mu\nabla_\mu\!\left(\nabla\!\!\cdot\!\vartheta\right)
\end{split}\\
&=\nabla_\mu\left[\vartheta_\nu\nabla^\mu\vartheta^\nu-
\vartheta^\mu\,\nabla\!\!\cdot\!\vartheta\right],
\end{align}
\end{subequations}
establishing finally that $\Delta W$ is a total divergence. Thus, it 
contributes nothing to the EoM and, unlike the G-B term, also zero from the 
boundary of a compact manifold.

For example, in four dimensions, the Lagrangian density $\lcal_{ho}$, 
\eqn{eq:hoeinstein2}, involving the real field $\phi(x)$ becomes
\begin{align}\label{eq:hoconf}
\begin{split}
\lcal_{ho}&{=}\sqrt{g} \bigg(\frac{1}{3b}\big[R {-} 6\,\Box\varUpsilon {+} 
6(\nabla\varUpsilon)^2 \big]^2+
 \frac{1}{2a} C_{\kappa\lambda\mu\nu}^2+\\
&\hskip10mm c R^*R^*\bigg),
 \end{split}
\end{align}
with $\varUpsilon\equiv(1/2)\log(\phi^2/M^2).$ 
The last term, which takes the form of a divergence locally, can be ignored in 
perturbation theory. With the form of $\lcal_{ho}$ in \eqn{eq:hoconf}, the
full action can then be written as
\begin{subequations}\label{eq:jactionconf}
\begin{align} 
\begin{split}
S_E&=\int d^4x \sqrt{g} \Big[-\frac{\xi M^2}{2}\big[R - 6\,\nabla^2\varUpsilon+\\
&\hskip10mm 6(\nabla\varUpsilon)^2 \big]+
\lcal_{ho}+\lcal_J(\phi,{g_\mn})\Big],
\end{split}\\
&\mathrm{where~} \lcal_J(\phi,{g_\mn})=\frac{ZM^2}{2}\big( 
{\nabla}\varUpsilon\big)^2{+}\frac{\lambda M^4}{4}\label{eq:jlagconf}.
\end{align}
\end{subequations}

The linear term in $R$ now has Einstein-Hilbert form, and the original $\phi^4$ 
self-interaction has become a cosmological constant that is positive for 
$\lambda{>}0\,$! Having assumed that $\phi(x)\ne0$, we may take 
$\phi>0,$ WLOG, since the action is invariant under $\phi\to-\phi.$ 

What remains is to gather like terms together in \eqn{eq:jactionconf}. The terms 
quadratic in $\varUpsilon$ are
\beq
\half\left(Z+6\xi \right)\left(M\nabla_\mu\varUpsilon\right)^2,
\eeq
where we temporarily neglected other terms coming from $\lcal_{ho}.$ Assuming 
that $Z+6\xi>0$, the canonically normalized scalar field is 
$\zeta\equiv\sqrt{(Z{+}6\xi)}\,M\varUpsilon.$ The preceding action then becomes 
\eqn{eq:eaction3} in the text.

We see that terms in $\lcal_{ho}$ involve powers of $\zeta/M.$ Similarly, if we 
carry out a derivative expansion in the metric as usual, then terms involving 
gradients of the metric beyond the quadratic terms and all those coming from 
$\lcal_{ho}$ will carry inverse powers of $M.$ Thus, while $\lcal_{ho}$ is critical 
for renormalizability, the low energy effective theory at energy scales small 
compared to $M$ will be dominated by the Einstein-Hilbert action as usual. Of 
course, $M$ is completely arbitrary here, but eventually in the QFT, we hope to 
reconcile this with the observed value.

\section{Lie algebra conventions}\label{sec:algebra}

We briefly review our conventions~\cite{Einhorn:2016mws} for the Lie algebra 
of $SO(10)$ in order to establish our notation and conventions. The defining or 
fundamental representation of the \emph{group} $SO(10)$ consists of 
$10{\times}10$ real, orthogonal matrices, $\ocal$ satisfying 
$\ocal\ocal^\tau=1,$ 
where $\ocal^\tau$ denotes the transpose. Writing $\ocal=\exp(i\theta_aR^a),$ 
the (Hermitian) generators $R^a$ must be imaginary and antisymmetric, 
satisfying 
\begin{align}\label{eq:so10basis}
\left[R^a,R^b\right]= if^{abc}R^c,\quad
\Tr [R^a R^b] =\delta^{ab}/2,
\end{align}
where we adopted the usual normalization convention (in physics) for the 
fundamental. Representation matrices are considered equivalent if they differ
only by a unitary transformation $\widetilde{\rcal}^a=U^\dag \rcal^a U.$ This is 
because the transformed matrices $\widetilde{\rcal}^a$ are Hermitian and still 
satisfy \eqn{eq:so10basis} with the same structure constants $f^{abc}.$ On the 
other hand, these equivalent matrices may be neither real nor antisymmetric.  In 
particular, it is possible to choose them so that the Cartan subalgebra is 
diagonal. (See, e.g.\ \reference{Cahn:1985wk}.) This is of considerable 
advantage for analyzing the patterns of SSB.

In order to understand better the choice of basis described in \eqn{eq:basis} 
and thereafter, we can proceed as follows. The $10{\times}10$ Hermitian 
generators $R^a$ are broken down into 4 $5{\times}5$ blocks of the form
given in \eqn{eq:basis}. The $\rcal_1^a$ are $5{\times}5$ Hermitian matrices,  of 
which there are 25 linearly independent possibilities. These 25 constitute a 
complete set that satisfy the algebra of $U(5){=}\su51,$ with the 
Cartan subalgebra given by the diagonal generators. We associate them with 
the first 25 $SO(10)$ generators, defining $\rcal_1^a=0$ for 
$\{a{=}26,\ldots,45\}\!:$ 
\beq
R_1^a{\equiv}
\frac{1}{\sqrt2}
\begin{pmatrix}
\ \rcal^a_1 &\vline& \ 0 \\
\hline
 0{}&\vline& - \rcal^a_1{}^*
\end{pmatrix}\!,\ 
\{a{=}1,\ldots,25\}.
\eeq
We choose the first 24 $\rcal_1^a$ to be traceless, generators of the 
$\mathbf{5}$ of $SU(5),$ with the 25th proportional to $\mathbf{1_5},$ the 
generator of the $U(1),$ normalized as required by $SO(10).$ This is 
frequently written as $\mathbf{5_{-2}}$. The 25 conjugate matrices 
$\{- \rcal^a_1{}^*\}$ of $\su51$ are generators for $\mathbf{\overline{5}_2}.$ 

On the other hand, we may also employ these generators to define the 
adjoint field, 
\beq\label{eq:phiadj}
\varPhi_1\equiv\sum_1^{25}\phi_a\rcal_1^a,
\eeq
with $\phi_a$ real.  The components of the matrix $\varPhi_1$ transform as 
the $\mathbf{24_0}{\oplus}\mathbf{1_0}$ representation of $\su51.$

The $\rcal_2^a$ are complex, antisymmetric, $5{\times}5$ matrices, of which there 
are 10 linearly independent that we shall call $\rcal^n.$ Because these are 
antisymmetric, we have $R^n{}^\dag=-R^n{}^*.$ 
From these, we may form two sets of $10{\times}10,$ Hermitian matrices,
\beq
\!\!R_2^{24{+}2n}{\equiv}\!\frac{1}{\sqrt2}\!\!
\begin{pmatrix}
\!0 \!\!&\vline&\! \rcal^n \\
\hline
\!-\rcal^n{}^* \!\!&\!\vline& 0
\end{pmatrix}\!\!, 
R_2^{25{+}2n}{\equiv}\!\frac{1}{\sqrt2}\!\!
\begin{pmatrix}
\!0 \!&\vline&\! i\rcal^n \\
\hline
i\rcal^n{}^* \!\!&\!\vline& 0
\end{pmatrix}\!\!,\!\!
\eeq
for $\{n{=}1,\ldots,10\}.$
Although the sub-blocks are obviously not linearly independent, the two 
sets are linearly independent as $SO(10)$ generators. We define 
$R_2^a=0$ for $\{a{=}1,\ldots,25\}.$

These too may be used to compose fields 
\beq
\varPhi_2\equiv\sum_1^{10}(\phi_{24{+}2n}{+}i\phi_{25{+}2n})\rcal^n\!,
\eeq
with real $(\phi_{24{+}2n}, \phi_{25{+}2n}).$ In fact, it can be shown that 
$(\varPhi_2)_{ij}$ transforms as the antisymmetric product representation 
$({\bf5_{-2}}{\otimes}{\bf5_{-2}})_a={\bf{\overline{10}}_{-4}}$ of 
the $U(5).$ Consequently, $-(\varPhi_2^*)_{ij}{\equiv} -(\varPhi_2)^{ij}$ 
transforms as the conjugate representation 
$({\bf{\overline{5}_2}}{\otimes}{\bf{\overline{5}}_2})_a={\bf{10}_{+4}}.$
(See, e.g.\ \reference{Slansky:1981yr}, Tables 29\,\&\,43.)

Combining these 45 component fields, we may write the adjoint of $SO(10)$ 
in the block form
\beq\label{eq:blockform}
\Phi=\sqrt2\phi_aR^a=
\begin{pmatrix}
\ \varPhi_1&\vline&\varPhi_2 \\
\hline
-\varPhi_2^* &\vline&\!\! -\varPhi_1^*
\!\end{pmatrix}.
\eeq

Indeed, the preceding decomposition describes the branching rules for 
$SO(10){\to}\su51,$ \viz
\beq \label{eq:branching}
\bf{45}\to\mathbf{1_0}\oplus\mathbf{24_0}\oplus
\mathbf{\overline{10}_{-4}}\oplus\mathbf{10_{+4}}.
\eeq
The first two $\mathbf{1_0}, \mathbf{24_0}$ are self-conjugate, whereas 
the last two are distinct conjugate pairs.
To break down this manner of representing $SO(10)$ into greater detail, since 
the $\rcal_1^a$ generate the algebra of $\su51,$ which has rank five, we can 
choose generators of the Cartan subalgebra, $\hcal_1^i, i{=}\{1,\ldots,5\},$  
to be diagonal. Setting $\hcal_2^i=0,$ the corresponding five $SO(10)$ 
generators are
\beq\label{eq:cartan}
H^i\equiv
\frac{1}{\sqrt2}
\begin{pmatrix}
\ \hcal^i_1 &\vline& \ 0 \\
\hline
 0{}&\vline& - \hcal^i_1{}^*
\end{pmatrix},\ 
\{i{=}1,\ldots,5\}.
\eeq
With the appropriate normalization of the $U(1)$ generator, we may assume 
$\Tr[H^i H^j]=\Tr[\hcal_1^i\hcal_1^j]=\delta^{ij}/2,$ as in \eqn{eq:so10basis}, \ie 
the $H^i$ are the Cartan generators of $SO(10)$ as well. In the text, this was
applied to the field $\vsig,$ decomposed as in \eqn{so10sigma}. It immediately 
follows that the expectation values obey \eqn{eq:vevCartan}.

Similarly, the real vector fields, $A^\mu,$ which transform as the adjoint of 
$SO(10),$ may be defined analogously to $\Phi,$ \eqn{eq:blockform},
$A^\mu\equiv\sqrt2 A_a^\mu R^a.$


\newpage
\begin{thebibliography}{99}

\bibitem{Einhorn:2014gfa}
 M.~B.~Einhorn and D.~R.~T.~Jones,
 ``Naturalness and Dimensional Transmutation in 
 Classically Scale-Invariant Gravity,''
 JHEP {\bf 1503} (2015) 047.
 %doi:10.1007/JHEP03(2015)047
 arXiv:1410.8513 [hep-th].
 
\bibitem{Jones:2015son}
 T.~Jones and M.~Einhorn,
 ``Quantum Gravity and Dimensional Transmutation,''
 PoS PLANCK {\bf 2015} (2015) 061.
 %%CITATION = POSCI,PLANCK2015,061;%%
 
\bibitem{Einhorn:2015lzy}
 M.~B.~Einhorn and D.~R.~T.~Jones,
 ``Induced Gravity I: Real Scalar Field,''
 JHEP {\bf 1601} (2016) 019.
 %doi:10.1007/JHEP01(2016)019
 arXiv:1511.01481 [hep-th].

\bibitem{Einhorn:2016mws}
 M.~B.~Einhorn and D.~R.~T.~Jones,
 ``Induced Gravity II: Grand Unification,''
 JHEP {\bf 1605} (2016) 185.
 %doi:10.1007/JHEP05(2016)185
arXiv:1602.06290 [hep-th].
 
 \bibitem{Einhorn:2016fmb}
 M.~B.~Einhorn and D.~R.~T.~Jones,
 ``Zero modes in de Sitter background,''
 JHEP {\bf 1703} (2017) 144.
 %doi:10.1007/JHEP03(2017)144
 arXiv:1606.02268 [hep-th].

\bibitem{Stelle:1976gc}
K.~S.~Stelle,
``Renormalization of Higher Derivative Quantum Gravity,''
Phys.\ Rev.\ D {\bf 16} (1977) 953.
%%CITATION = PHRVA,D16,953;%%

\bibitem{Fradkin:1981hx}
E.~S.~Fradkin and A.~A.~Tseytlin,
``Renormalizable Asymptotically Free 
Quantum Theory of Gravity,''
Phys.\ Lett.\ {\bf 104B} (1981) 377;
 %%CITATION = PHLTA,B104,377;
%\bibitem{Fradkin:1981iu}

 %E.~S.~Fradkin and A.~A.~Tseytlin,
 {\it id.}, ``Renormalizable asymptotically free 
quantum theory of gravity,''
 Nucl.\ Phys.\ B {\bf 201} (1982) 469.
 %%CITATION = NUPHA,B201,469;%%
 
\bibitem{Giudice:2014tma}
  G.~F.~Giudice, G.~Isidori, A.~Salvio and A.~Strumia,
``Softened Gravity and the Extension of the Standard Model up to Infinite Energy,''
  JHEP {\bf 1502} (2015) 137.
  %doi:10.1007/JHEP02(2015)137
  arXiv:1412.2769 [hep-ph]. 
 
\bibitem{Weinberg:1976xy}
 S.~Weinberg,
 ``Critical Phenomena for Field Theorists,'' in
``Understanding the Fundamental Constituents of Matter: proceedings, Edited 
by A.Zichichi~(ed.), N.Y., Plenum Press, 1978. 915p. (The Subnuclear Series, 
\#14). 

\bibitem{Litim:2014uca}
  D.~F.~Litim and F.~Sannino,
 ``Asymptotic safety guaranteed,''
  JHEP {\bf 1412} (2014) 178
  doi:10.1007/JHEP12(2014)178.
  arXiv:1406.2337 [hep-th];
  
%\bibitem{Bond:2017wut}
 A.~D.~Bond {\it et al.},
 ``Directions for model building from asymptotic safety,''
arXiv:1702.01727 [hep-ph];

%\bibitem{Reuter:2012id}
 M.~Reuter and F.~Saueressig,
 ``Quantum Einstein Gravity,''
 New J.\ Phys.\ {\bf 14} (2012) 055022;
 %doi:10.1088/1367-2630/14/5/055022
 arXiv:1202.2274 [hep-th].

%\bibitem{Demmel:2014hla}
 M.~Demmel, F.~Saueressig and O.~Zanusso,
 ``RG flows of Quantum Einstein Gravity in the linear-geometric approximation,''
 Annals Phys.\ {\bf 359} (2015) 141.
 %doi:10.1016/j.aop.2015.04.018
 arXiv:1412.7207 [hep-th].
 
\bibitem{Bardeen:1995kv}
W.~A.~Bardeen,
``On naturalness in the standard model,''
Fermilab-Conf-95-391-T, (unpublished)
%%CITATION = FERMILAB-CONF-95-391-T;%%
and W.~A.~Bardeen, private communication. 

\bibitem{Coleman:1973jx}
S.~R.~Coleman and E.~J.~Weinberg,
``Radiative Corrections as the Origin of Spontaneous Symmetry Breaking,''
Phys.\ Rev.\ D {\bf 7} (1973) 1888.
%%CITATION = PHRVA,D7,1888;%%

\bibitem{Avramidi:1986mj}
 I.~G.~Avramidi,
 ``Covariant methods for the calculation of the effective action in quantum field theory and investigation of higher derivative quantum gravity,''
 hep-th/9510140;
 %%CITATION = HEP-TH/9510140;%%
 
%\bibitem{Avramidi:2000bm}
 %I.~G.~Avramidi,
{\it id.},
``Heat kernel and quantum gravity,''
 Lect.\ Notes Phys.\ Monogr.\ {\bf 64} (2000) 1.
 %doi:10.1007/3-540-46523-5
 
\bibitem{Barth:1983hb}
 N.~H.~Barth and S.~M.~Christensen,
 ``Quantizing Fourth Order Gravity Theories. 1. The Functional Integral,''
 Phys.\ Rev.\ D {\bf 28} (1983) 1876.
 %doi:10.1103/PhysRevD.28.1876 
 
\bibitem{Holdom:2016xfn}
 B.~Holdom and J.~Ren,
 ``Quadratic gravity: from weak to strong,''
 Int.\ J.\ Mod.\ Phys.\ D {\bf 25} (2016) no.12, 1643004.
 %doi:10.1142/S0218271816430045
 arXiv:1605.05006 [hep-th].
 
\bibitem{Fradkin:1983tg}
 E.~S.~Fradkin and A.~A.~Tseytlin,
 ``Conformal Anomaly in Weyl Theory and Anomaly Free Superconformal Theories,''
 Phys.\ Lett.\ {\bf 134B} (1984) 187.
 %doi:10.1016/0370-2693(84)90668-3 
 
\bibitem{Fradkin:1985am}
 E.~S.~Fradkin and A.~A.~Tseytlin,
 ``Conformal Supergravity,''
 Phys.\ Rept.\ {\bf 119} (1985) 233.
 %doi:10.1016/0370-1573(85)90138-3 

\bibitem{Adler:1982ri}
 S.~L.~Adler,
 ``Einstein Gravity as a Symmetry Breaking Effect in Quantum Field Theory,''
 Rev.\ Mod.\ Phys.\ {\bf 54} (1982) 729.
 Erratum: [Rev.\ Mod.\ Phys.\ {\bf 55} (1983) 837]; 
 %doi:10.1103/RevModPhys.54.729
 
%\bibitem{Buchbinder:1986wk}
 I.~L.~Buchbinder,
 ``Mechanism For Induction Of Einstein Gravitation,''
 Sov.\ Phys.\ J.\ {\bf 29} (1986) 220;
 %doi:10.1007/BF00891883
 %%CITATION = %doi:10.1007/BF00891883;%%
 
% \bibitem{Odintsov:1991nd}
  S.~D.~Odintsov and I.~L.~Shapiro,
  ``General relativity as the low-energy limit in higher derivative quantum gravity,''
  Class.\ Quant.\ Grav.\  {\bf 9} (1992) 873
   [Theor.\ Math.\ Phys.\  {\bf 90} (1992) 319];
   %[Teor.\ Mat.\ Fiz.\  {\bf 90} (1992) 469].
  %doi:10.1088/0264-9381/9/4/006, 10.1007/BF01036537
  
%\bibitem{Shapiro:1994yt}
 I.~L.~Shapiro,
 ``Hilbert-Einstein action from induced gravity coupled with scalar field,''
 Mod.\ Phys.\ Lett.\ A {\bf 9} (1994) 1985.
 %doi:10.1142/S0217732394001842
 [hep-th/9403077].; 
 
%\bibitem{Cognola:1998ve}
 G.~Cognola and I.~L.~Shapiro,
``Back reaction of vacuum and the renormalization group flow from the conformal fixed point,''
 Class.\ Quant.\ Grav.\ {\bf 15} (1998) 3411;
 %doi:10.1088/0264-9381/15/11/008
 %[hep-th/9804119].
 %%CITATION = %doi:10.1088/0264-9381/15/11/008;%% 
 
%\bibitem{Kannike:2015apa}
 K.~Kannike et al,
``Dynamically Induced Planck Scale and Inflation,''
 JHEP {\bf 1505} (2015) 065.
 %doi:10.1007/JHEP05(2015)065
 arXiv:1502.01334 [astro-ph.CO].
 
\bibitem{Buchbinder:1992rb}
I.~L.~Buchbinder, S.~D.~Odintsov and I.~L.~Shapiro,
``Effective action in quantum gravity,''
Bristol, UK: IOP (1992).  

\bibitem{Salvio:2014soa}
 A.~Salvio and A.~Strumia,
``Agravity,''
JHEP {\bf 1406} (2014) 080.
arXiv:1403.4226 [hep-ph].

\bibitem{Salvio:2017qkx}
  A.~Salvio and A.~Strumia,
  ``Agravity up to infinite energy,''
  Eur.\ Phys.\ J.\ C {\bf 78} (2018) no.2,  124.
  %doi:10.1140/epjc/s10052-018-5588-4
  arXiv:1705.03896 [hep-th].



\bibitem{GarciaBellido:2011de}
  J.~Garcia-Bellido, {\it et al.},
  %J.~Rubio, M.~Shaposhnikov and D.~Zenhausern,
 ``Higgs-Dilaton Cosmology: From the Early to the Late Universe,''
  Phys.\ Rev.\ D {\bf 84} (2011) 123504;
  %doi:10.1103/PhysRevD.84.123504
  arXiv:1107.2163 [hep-ph];
  
%\bibitem{Ferreira:2016vsc}
 P.~G.~Ferreira, C.~T.~Hill and G.~G.~Ross,
 ``Scale-Independent Inflation and Hierarchy Generation,''
 Phys.\ Lett.\ {\bf 763B} (2016) 174.
 %doi:10.1016/j.physletb.2016.10.036
 arXiv:1603.05983 [hep-th].
 
\bibitem{Hawking:1978jn}
 S.~W.~Hawking,
 ``Euclidean Quantum Gravity,'' lectures presented at the 
 %Cargese\ Summer\ Inst.\ 1978, 
 NATO Sci.\ Ser.\ B {\bf 44} (1979) 145; 
 (included in 
 % \bibitem{Levy:1979im}
 M.~Levy and S.~Deser,
 ``Recent Developments In Gravitation. Proceedings, Summer Institute, 
 Cargese, Corsica, July 10-29, 1978,'' N.Y.: Plenum Press, (1979));
 
%\bibitem{Gibbons:1994cg}
 G.~W.~Gibbons and S.~W.~Hawking, (eds.)
 ``Euclidean quantum gravity,''
 Singapore: World Scientific (1993) 586 p.;
 
%\bibitem{DeWitt:1978gi}
 B.~S.~DeWitt,
 ``The Formal Structure of Quantum Gravity,'' lectures presented at the 
 Cargese\ Summer\ Inst.\ 1978, 
 NATO Sci.\ Ser.\ B {\bf 44} (1979) 275; included in Levy \& Deser, 
 {\it op.cit.,}
% \reference{Levy:1979im} 
 and reprinted in %\reference{DeWitt:2007mi};
 %%CITATION = INSPIRE-138682;%%
% \bibitem{DeWitt:2007mi}
 B.~S.~DeWitt and G.~Esposito,
 ``An Introduction to quantum gravity,''
 Int.\ J.\ Geom.\ Meth.\ Mod.\ Phys.\ {\bf 5} (2008) 101. 
 %doi:10.1142/S0219887808002679
 arXiv:0711.2445 [hep-th].

\bibitem{Christensen:1979iy}
 S.~M.~Christensen and M.~J.~Duff,
 ``Quantizing Gravity with a Cosmological Constant,''
 Nucl.\ Phys.\ B {\bf 170} (1980) 480.
 %doi:10.1016/0550-3213(80)90423-X

\bibitem{Osterwalder:1973dx}
 K.~Osterwalder and R.~Schrader,
 ``Axioms For Euclidean Green's Functions,''
 Commun.\ Math.\ Phys.\ {\bf 31} (1973) 83;
 %doi:10.1007/BF01645738
 
%\bibitem{Osterwalder:1974tc}
 %K.~Osterwalder and R.~Schrader,
 {\it id.},
 ``Axioms for Euclidean Green's Functions. 2.,''
 Commun.\ Math.\ Phys.\ {\bf 42} (1975) 281.
 %doi:10.1007/BF01608978

\bibitem{Sotiriou:2008rp}
 T.~P.~Sotiriou and V.~Faraoni,
 ``f(R) Theories Of Gravity,''
 Rev.\ Mod.\ Phys.\ {\bf 82} (2010) 451;
 %doi:10.1103/RevModPhys.82.451
 arXiv:0805.1726 [gr-qc].
 
%\bibitem{DeFelice:2010aj}
 A.~De Felice and S.~Tsujikawa,
 ``f(R) theories,''
 Living Rev.\ Rel.\ {\bf 13} (2010) 3;
 %doi:10.12942/lrr-2010-3
arXiv:1002.4928 [gr-qc].

%\bibitem{Nojiri:2010wj}
  S.~Nojiri and S.~D.~Odintsov,
  ``Unified cosmic history in modified gravity: from F(R) theory to Lorentz non-invariant models,''
  Phys.\ Rept.\  {\bf 505} (2011) 59;
 % doi:10.1016/j.physrep.2011.04.001
arXiv:1011.0544 [gr-qc].

%\bibitem{Narain:2016sgk}
 G.~Narain,
 ``Exorcising Ghosts in Induced Gravity,''
 arXiv:1612.04930 [hep-th].
 %%CITATION = ARXIV:1612.04930;%%

\bibitem{Jack:2000nm} 
 I.~Jack, D.~R.~T.~Jones and S.~Parsons,
 ``The Fayet-Iliopoulos D term and its renormalization 
in softly broken supersymmetric theories,''
 Phys.\ Rev.\ D {\bf 62}, 125022 (2000).
 %doi:10.1103/PhysRevD.62.125022
 [hep-ph/0007291].

\bibitem{Rodrigues:2011zi}
 D.~C.~Rodrigues, {\it et al.},
% F.~de O.Salles, I.~L.~Shapiro and A.~A.~Starobinsky,
 ``Auxiliary fields representation for modified gravity models,''
 Phys.\ Rev.\ D {\bf 83} (2011) 084028.
 %doi:10.1103/PhysRevD.83.084028
 arXiv:1101.5028 [gr-qc]. 
 
\bibitem{Narain:2017tvp}
  G.~Narain,
  %``Signs and Stability in Higher-Derivative Gravity,''
  Int.\ J.\ Mod.\ Phys.\ A {\bf 33} (2018) no.04,  1850031.
  %doi:10.1142/S0217751X18500318
  arXiv:1704.05031 [hep-th].

 
 
 \bibitem{Birrell:1982ix}
  N.~D.~Birrell and P.~C.~W.~Davies,
  ``Quantum Fields in Curved Space,''
  Cambridge Monographs on Mathematical Physics,
  Cambridge Univ. Press, 1982.
  %doi:10.1017/CBO9780511622632
 
\bibitem{Parker:2009uva}
  L.~E.~Parker and D.~Toms,
  ``Quantum Field Theory in Curved Spacetime : Quantized Field and Gravity,''
  Cambridge University Press, 2009.
 %%doi:10.1017/CBO9780511813924;%%  

\bibitem{Nelson:2010ig}
 W.~Nelson,
 ``Static Solutions for 4th order gravity,''
 Phys.\ Rev.\ D {\bf 82} (2010) 104026.
 %doi:10.1103/PhysRevD.82.104026
 arXiv:1010.3986 [gr-qc].
 
\bibitem{Hamermesh:grpth}
M.~Hamermesh,
``Group Theory and Its Application to Physical Problems,"
Addison-Wesley (1962), reprinted by Dover Books on Physics,
Dover Pubs.~(1989). 

\bibitem{Hawking:1976ja}
 S.~W.~Hawking,
 ``Zeta Function Regularization of Path Integrals in Curved Space-Time,''
 Commun.\ Math.\ Phys.\ {\bf 55} (1977) 133.
 %doi:10.1007/BF01626516

\bibitem{Elizalde:1997nd}
 E.~Elizalde, L.~Vanzo and S.~Zerbini,
 ``Zeta function regularization, the multiplicative anomaly and the Wodzicki residue,''
 Commun.\ Math.\ Phys.\ {\bf 194} (1998) 613.
 %doi:10.1007/s002200050371
[hep-th/9701060].

\bibitem{Flanagan:2004bz}
 E.~E.~Flanagan,
 ``The Conformal frame freedom in theories of gravitation,''
 Class.\ Quant.\ Grav.\ {\bf 21} (2004) 3817.
 %doi:10.1088/0264-9381/21/15/N02
 [gr-qc/0403063].
 %%CITATION = doi:10.1088/0264-9381/21/15/N02;%%

 \bibitem{tHooft:1976snw}
 G.~'t Hooft,
 ``Computation of the Quantum Effects Due to a Four-Dimensional Pseudoparticle,''
 Phys.\ Rev.\ D {\bf 14} (1976) 3432; 
 Erratum: [Phys.\ Rev.\ D {\bf 18} (1978) 2199].
 % doi:10.1103/PhysRevD.18.2199.3, 10.1103/PhysRevD.14.3432.

 \bibitem{'tHooft:1973mm}
 G.~'t Hooft,
``Dimensional regularization and the renormalization group,''
 Nucl.\ Phys.\ B {\bf 61} (1973) 455.
 %%CITATION = NUPHA,B61,455;%% 

\bibitem{Alvarez-Gaume:2015rwa}
  L.~Alvarez-Gaume, {\it et al.},
  %A.~Kehagias, C.~Kounnas, D.~L\"ust, and A.~Riotto,
  ``Aspects of Quadratic~Gravity,''
  Fortsch.\ Phys.\  {\bf 64} (2016) no.~2-3,  176.
 % doi:10.1002/prop.201500100
  arXiv:1505.07657 [hep-th].

\bibitem{Carroll:2004st}
 S.~M.~Carroll,
 ``Spacetime and geometry: An introduction to general relativity,''
 San Francisco, USA: Addison-Wesley (2004) 513 p. 
  
\bibitem{AlvarezGaume:1984dr}
 L.~Alvarez-Gaume and P.~H.~Ginsparg,
 ``The Structure of Gauge and Gravitational Anomalies,''
 Annals Phys.\ {\bf 161} (1985) 423. 
 [Erratum-ibid.\ {\bf 171} (1986) 233].
 %%CITATION = APNYA,161,423;%%

\bibitem{Cahn:1985wk}
 R.~N.~Cahn,
 ``Semisimple Lie Algebras And Their Representations,''
 Menlo Park, USA: Benjamin/Cummings (1984) 158p. 
(Frontiers In Physics, \#59).

\bibitem{Slansky:1981yr}
 R.~Slansky,
 ``Group Theory for Unified Model Building,''
 Phys.\ Rept.\ {\bf 79} (1981) 1.
 %doi:10.1016/0370-1573(81)90092-2

\end{thebibliography}
\end{document}